\documentclass[iop]{emulateapj}

\usepackage[colorlinks,citecolor=blue, linkcolor=blue]{hyperref}
\usepackage{graphicx}
\usepackage{color,soul}
\usepackage{natbib}
\usepackage{amsmath}
\usepackage{url}

\def \msun{$M_{\odot}$\ }
\def \ovii{\ion{O}{7}\ }
\def \oviii{\ion{O}{8}\ }
\defcitealias{miller_bregman15}{MB15}
\defcitealias{kataoka_etal15}{K15}
\defcitealias{hs12}{HS12}

\begin{document}

\title{The Interaction of the Fermi Bubbles with the Milky Way's Hot Gas Halo}
\author{Matthew J. Miller \& Joel N. Bregman }
\affil{Department of Astronomy, University of Michigan, Ann Arbor, MI 48104, USA}
\email{mjmil@umich.edu, jbregman@umich.edu}

\begin{abstract}

The Fermi bubbles are two lobes filled with non-thermal particles that emit gamma rays, extend $\approx$10 kpc vertically from the Galactic center, and formed from either nuclear star formation or accretion activity on Sgr A*.  Simulations predict a range of shock strengths as the bubbles expand into the surrounding hot gas halo distribution ($T_{halo} \approx 2 \times 10^6$ K), but with significant uncertainties in the energetics, age, and thermal gas structure.  The bubbles should contain thermal gas with temperatures between $10^6$ and $10^8$ K, with potential X-ray signatures.  In this work, we constrain the bubbles' thermal gas structure by modeling the \ion{O}{7} and \ion{O}{8} emission line strengths from archival \textit{XMM-Newton} and \textit{Suzaku} data.  Our emission model includes a hot thermal volume-filled bubble component cospatial with the gamma-ray region, and a shell of compressed material.  We find that a bubble/shell model with $n \approx 1 \times 10^{-3}$ cm$^{-3}$ and with log($T$) $\approx$ 6.60-6.70 is consistent with the observed line intensities.  In the framework of a continuous Galactic outflow, we infer a bubble expansion rate, age, and energy injection rate of $490_{-77}^{+230}$ km s$^{-1}$, $4.3_{-1.4}^{+0.8}$ Myr, and $2.3_{-0.9}^{+5.1} \times 10^{42}$ erg s$^{-1}$.  These estimates are consistent with the bubbles forming from a Sgr A* accretion event rather than from nuclear star formation.

\end{abstract}

\keywords{Galaxy: halo --- Galaxy: center --- X-rays: diffuse background --- X-rays: ISM}

\section{Introduction}
\label{section.introduction_chap_fb}

The Fermi bubbles are important Galactic structures that were recently discovered by the \textit{Fermi Gamma-ray Space Telescope} \citep{su_etal10}.  The bubbles are two diffuse lobes of material extending $\sim$50$\arcdeg$ above and below the Galactic plane ($\approx$10 kpc at the Galactic center).  Their surface brightness shows little variation on the sky, their gamma-ray spectrum follows a power law with $dN/dE \propto E^{-2}$ between $\approx$1 and 200 GeV, and they have a counterpart in microwaves, known as the $Wilkinson\ Microwave\ Anisotropy\ Probe$ ($WMAP$) haze \citep{dobler_finkbeiner08, dobler_etal10, su_etal10, ackermann_etal14}.  It is still unclear what produces the gamma rays, but all plausible mechanisms imply that energetic cosmic-ray particles exist within the bubbles.  This inference combined with the bubbles' size and location on the sky suggests that they are affiliated with a massive energy injection event near the Galactic center.  

The bubbles' morphology is similar to wind-blown bubbles observed in other galaxies, indicating that they formed from either a period of enhanced nuclear star formation or a Sgr A* outburst event (see \citealt{veilleux_etal05} for a review).  Star formation can drive outflows through a combination of stellar winds from young stars and multiple type-II supernova explosions \citep[e.g., ][]{leitherer_etal99}, while black hole accretion episodes can produce energetic jets or winds that inflate a cavity with thermal and non-thermal particles \citep[e.g., ][]{mcnamara_nulsen07,yuan_narayan14}.  Both of these scenarios are critical events in galaxy evolution, as they both can deposit significant amounts of energy into the rest of the galaxy on scales $\gtrsim$10 kpc (see \citet{mcnamara_nulsen07} for a review).  However, the details of these ``feedback'' effects (mass displacement, energy transport, etc.) are poorly understood since we observe them in external galaxies.  The Fermi bubbles are a unique laboratory for understanding these processes since we can spatially resolve the bubbles across multiple wavebands.  

A popular strategy to probe these effects and bubbles' origins has been the use of magnetohydrodynamic (MHD) simulations to reproduce the bubbles’ global morphology and non-thermal properties.  Simulations produce cosmic rays either from a black hole accretion event \citep{zubovas_etal11,guo_mathews12a,guo_mathews12b,yang_etal12,yang_etal13,zubovas_nayakshin12,mou_etal14,mou_etal15}, from nuclear star formation activity \citep{crocker12,crocker_etal14,crocker_etal15,sarkar_etal15,ruszkowski_etal16}, or in-situ as the bubbles evolve \citep{cheng_etal11,cheng_etal15b,mertsch_sarkar11,fujita_etal14,lacki14,sasaki_etal15}, and compare the non-thermal emission to the bubbles' gamma-ray emission.  All of these origin scenarios can reproduce the bubbles' morphology, but they imply significantly different input energetics and timescales required to inflate the bubbles ($\dot{E} \gtrsim 10^{41}$ erg s$^{-1}$, $t \lesssim 5$ Myr for black hole accretion compared to $\dot{E} \lesssim 5 \times 10^{40}$ erg s$^{-1}$, $t \gtrsim 50$ Myr for star formation).  This variation in the feedback rate is a significant uncertainty in how the bubbles impact the Galaxy, but there are additional factors that can constrain the characteristic bubble energetics.  

Constraining the bubbles' thermal gas distribution is a promising avenue to solve this problem, since the characteristic densities and temperatures should be significantly different depending on the bubble energetics.  In the framework of expanding galactic outflows and shocks \citep[e.g., ][]{veilleux_etal05}, a higher energy input rate leads to a higher plasma temperature and a larger expansion rate for a fixed bubble size and ambient density.  Thus, the plasma temperature at the interface between the bubbles and surrounding medium encodes information on the bubbles' shock strength, expansion properties, and overall energy input rate.  A generic prediction from simulations and observations of galactic outflows is that the bubbles are overpressurized and hotter than the surrounding medium ($\gtrsim 2 \times 10^6$ K), implying that the bubbles' thermal gas should have signatures at soft X-ray energies.  Indeed, the bubbles appear to be bounded by X-ray emission seen in the \textit{ROSAT} 1.5 keV band \citep{bh_cohen03, su_etal10}; however, these observations do not constrain the bubbles' intrinsic thermal gas structure since the broad-band images are a weak temperature diagnostic.  Spectral observations with current X-ray telescopes are a much better temperature diagnostic for this type of environment.  

Initial efforts to observe the bubbles in soft X-rays with \textit{Suzaku} and \textit{Swift} and constrain their temperature and shock strength were carried out by several groups \citep{kataoka_etal13,kataoka_etal15,tahara_etal15}.  \citet{kataoka_etal15} extracted soft X-ray background (SXRB) spectra in the 0.5--2.0 keV band for 97 sight lines that pass through the Fermi bubbles, and fit the spectra with thermal plasma models.  They consistently measured plasma temperatures of $kT$ = 0.3 keV for these sight lines, which is systematically higher than the characteristic temperature measured in sight lines away from the Galactic center \citep[$kT \approx$0.2 keV; ][]{hs13}.  From this temperature ratio, they inferred a shock Mach number of $\mathcal{M} \approx $0.3 keV / 0.2 keV = 1.5, and corresponding expansion rate of $\approx$300 km s$^{-1}$.  This is a valuable attempt to constrain these quantities, but the analysis assumes that the Fermi bubble plasma dominates the hotter spectral component.  In practice, there are other known emission sources that contribute to the SXRB spectrum, and accounting for this emission can change the inferred thermal gas temperature.

The Milky Way hosts a hot gas distribution with $T \approx 2 \times 10^6$ K extending on scales $\gtrsim$10 kpc based on shadowing experiments from \textit{ROSAT} all-sky data \citep{snowden_etal97, kuntz_snowden00}.  This plasma is believed to dominate any SXRB spectrum, with \ovii and \oviii being the characteristic observed line transitions \citep[e.g., ][]{mccammon_etal02, yoshino_etal09, hs12}.  The structure of this extended plasma distribution has been debated in the literature, but numerous studies on but numerous studies on both absorption and emission line strengths indicate that the plasma is spherical and extends to at least $r \sim 50$ kpc \citep{fang_etal06, bregman_ld07, gupta_etal12, fang_etal13, miller_bregman13, miller_bregman15}.  In particular, Miller \& Bregman (\citeyear{miller_bregman15}, defined as \citetalias{miller_bregman15} henceforth) modeled a set of 648 \oviii emission line intensities from Henley \& Shelton (\citeyear{hs12}, defined as \citetalias{hs12} henceforth), and found that a hot gas density profile with $n \propto r^{-3/2}$ extending to the virial radius reproduces the observed emission line intensities.  These modeling studies have placed useful constraints on the Galactic-scale hot gas distribution, but also highlight the fact that this extended plasma is likely the dominant emission source in all 0.5--2.0 keV band spectra.  

In this study, we expand the analysis Kataoka et al. (\citeyear{kataoka_etal15}, defined as \citetalias{kataoka_etal15} henceforth) analysis by modeling the combined emission from the Fermi bubbles and hot gas halo present in \oviii emission line measurements.  We modify the Galactic-scale hot gas models from \citetalias{miller_bregman15} to include a geometry, density, and temperature structure for the Fermi bubbles.  Given a set of model parameters, we predict the contribution to the emission made by the Fermi bubbles and hot gas halo along any sight line.  This results in a more careful comparison between the Fermi bubbles' emission and the total observed emission in any SXRB measurement.

The \oviii observations used in our analysis consist of published \textit{XMM-Newton} measurements from \citetalias{hs12}, and a new \textit{Suzaku} data set produced for this work.  The \textit{XMM-Newton} data are mostly the same measurements used in \citetalias{miller_bregman15}, but we now include data near the Fermi bubbles.  We supplement these data with archival \textit{Suzaku} measurements of SXRB spectra, which more than doubles the number of emission line measurements projected near the bubbles.  These data are processed in a similar way to the \textit{XMM-Newton} data reduction outlined in \citetalias{hs12}, resulting in a uniformly processed data set of emission line intensities from the SXRB.  

Following the methodology from \citetalias{miller_bregman15}, we constrain the Fermi bubbles' density and temperature structure by finding the parametric model that is most consistent with the observed emission line intensities.  We measure the characteristic bubble temperature from analyzing the distribution of observed \oviii/\ovii line ratios near the bubbles, and the characteristic bubble density from explicitly modeling the \oviii emission line intensities.  We infer a similar bubble shock strength compared to the \citetalias{kataoka_etal15} analysis, and discuss the systematic differences between our approaches and results and theirs.  We also estimate the bubbles' age and energy input rate, and these with the possible formation mechanisms discussed above.

The rest of the paper is outlined as follows.  Section~\ref{section.data_chap_fb} discusses how we compiled our emission line sample, including an overview of the \textit{XMM-Newton} data set and the \textit{Suzaku} data processing.  Section~\ref{section.model_chap_fb} definesdefines our parametric density and temperature model and discusses our line intensity calculation.  Section~\ref{section.results_chap_fb} discusses our model fitting routine and results.  Section~\ref{section.discussion_chap_fb} discusses our constraints on the Fermi bubbles in the context of galactic outflows, previous X-ray studies, and simulations.  Section~\ref{section.conclusions_chap_fb} summarizes our results.  

\section{Emission Line Data}
\label{section.data_chap_fb}

Our data set includes \ovii (He-like triplet at $E \approx 0.56$ keV) and \oviii (Ly$\alpha$ transition at $E \approx 0.65$ keV) emission lines, which are the dominant ions for thermal plasmas with temperatures between $T \sim 10^{5.5}$ and $10^{7}$ K \citep{sutherland_dopita93}.  For an optically thin plasma in collisional ionization equilibrium, the emission line intensity depends on the plasma density and temperature as $I\ \propto\ n^2 \epsilon(T)$, where $\epsilon(T)$ is the volumetric line emissivity.  This implies that the line strength ratio is a temperature diagnostic and the individual ion line strengths can be used to estimate the plasma density.  Large, all-sky samples of emission line measurements in particular have been instrumental in constraining the Milky Way's global density distribution of $\sim 10^6$ K gas \citepalias{miller_bregman15}.  

The full data set used in our modeling analysis is a combination of published \textit{XMM-Newton} emission line measurements from \citetalias{hs12} and a complementary sample of \textit{Suzaku} measurements compiled specifically for this project.  The \textit{XMM-Newton} sample from \citetalias{hs12} contains $\sim$1000 emission line measurements distributed across the sky, making it a valuable starting point for our modeling work.  While \textit{XMM-Newton} has more collecting power than \textit{Suzaku} near the emission lines (collecting area $\times$ field of view $\approx$140,000 cm$^{2}$ arcmin$^{2}$ for the MOS1 camera compared to $\approx$70,000 cm$^{2}$ arcmin$^{2}$ for the XIS1 detector), we include a supplemental \textit{Suzaku} data set for two reasons.  \textit{Suzaku's} low Earth orbit results in a lower and more stable particle background than \textit{XMM-Newton's}, often resulting in measurements with higher signal-to-noise ratio at soft X-ray energies.  Also, there are many valuable archival \textit{Suzaku} observations projected near the Fermi bubbles, including 14 observations dedicated to observing the bubbles' edge.  In practice, \textit{Suzaku} data should have a comparable signal-to-noise ratio to the \textit{XMM-Newton} data, while probing the crucial region in and near the Fermi bubbles.  

Our goal is to create a clean sample of uniformly processed emission line measurements by reducing the \textit{Suzaku} data in a similar way to how \citetalias{hs12} reduced the \textit{XMM-Newton} data.  The main steps include: the removal of bright point sources, light curve filtering, spectral fitting, and filtering for solar wind charge-exchange (SWCX) emission.  The following sections summarize how the \textit{XMM-Newton} data were produced, and detail our \textit{Suzaku} data reduction steps.  After applying all these data reduction methods, our final sample includes 683 useful \textit{XMM-Newton} measurements and 58 useful \textit{Suzaku} measurements.

\subsection{\textit{XMM-Newton} Observations}
\label{subsection.xmm_chap_fb}

We summarize the \textit{XMM-Newton} emission line sample compilation here, but we refer the reader to \citetalias{hs12} for a more detailed description of their data reduction methods.  Their initial sample included 5698 observations that had any EPIC-MOS exposure time.  They reduced the data using the \textit{XMM-Newton} Extended Source Analysis Software \footnote{\url{http://heasarc.gsfc.nasa.gov/docs/xmm/xmmhp_xmmesas.html}} \citep[\textit{XMM}-ESAS;][]{kuntz_snowden08, snowden_kuntz11}, which includes screening the 2.5--12 keV band count rate for soft proton flares.  They also removed point sources from the spectral extraction regions using data from the Second \textit{XMM-Newton} Serendipitous Source Catalog \footnote{\url{http://xmmssc-www.star.le.ac.uk/Catalogue/2XMMi-DR3/}} \citep{watson_etal09}, as well as visual inspection for bright sources in the images.  The authors attempted to reduce geocoronal SWCX emission by excluding observing periods with high solar wind proton flux measurements (see Section~\ref{subsection.swcx_chap_fb} for the details of this procedure), which they called their ``flux-filtered'' sample.  These reduction methods resulted in 1868 total observations and 1003 flux-filtered observations with $\geq$5 ks of good observing time.  Each of these observations includes an \ovii and \oviii emission line intensity measurement, and they have been used to analyze the known emission sources (i.e., Local Bubble (LB), extended hot halo, SWCX).  

This sample has been used before to successfully model the Milky Way's global hot gas structure, thus motivating its use to model the Fermi bubbles.  \citetalias{miller_bregman15} compiled a subset of the \citetalias{hs12} flux-filtered sample with additional spatial screening criteria to reduce any residual emission from sources other than the LB and Galactic hot halo.  To achieve this, they removed observations within 0.5$\arcdeg$ of potential bright X-ray sources (see Table 1 in \citetalias{miller_bregman15} for the types of sources) or within 10$\arcdeg$ of the Galactic plane, and sight lines near the Fermi bubbles ($|l| \leq\ 22\arcdeg$, $|b| \leq\ 55\arcdeg$).  This resulted in a subsample of 649 observations from the \citetalias{hs12} flux-filtered sample.  

The \textit{XMM-Newton} data used in this study are the same as those used in \citetalias{miller_bregman15}, but including the sight lines near the Fermi bubbles.  We start with the \citetalias{hs12} flux-filtered sample and still exclude sight lines near bright X-ray sources or within 10$\arcdeg$ of the Galactic plane.  These screening criteria result in a total of 683 \textit{XMM-Newton} measurements distributed across the sky, with 34 measurements passing near or through the bubbles' gamma-ray edge.  

\subsection{\textit{Suzaku} Observations and Data Reduction}
\label{subsection.suz_chap_fb}


\begin{figure*}
\begin{center}
\includegraphics[width = 1.0\textwidth, keepaspectratio=true]{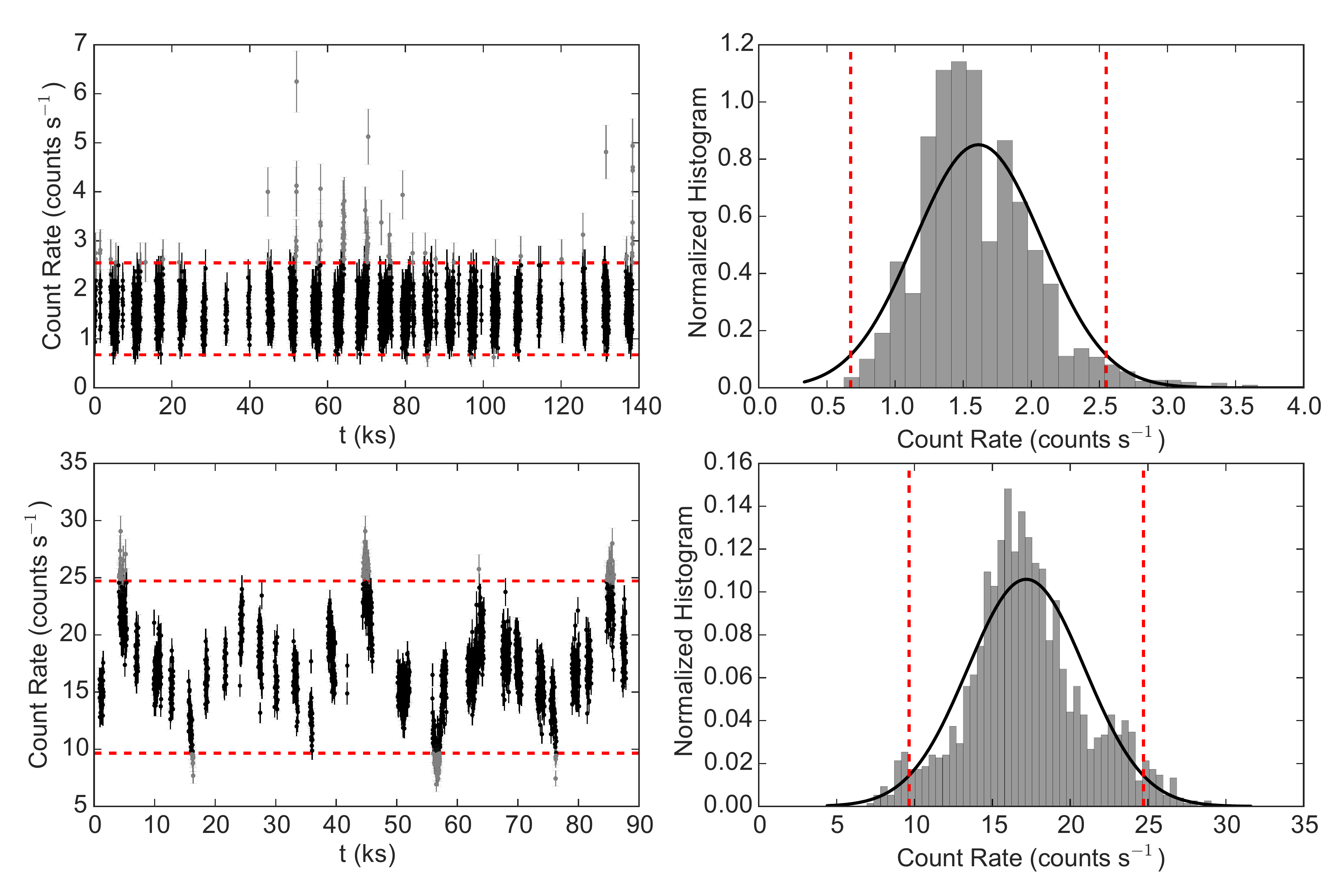}
\caption[Example light curve and count rate histogram]{Example 0.4--10.0 keV light curves (left panels) and count-rate histograms (right panels) with fitted Gaussian distributions for two observations in our initial sample.  Observing periods within the 2$\sigma$ limits (red dashed lines) were kept while periods outside of these limits were excluded (gray points in the left panels).  The top panels show a Galactic bulge observation (Obs. ID 100011010) with a well-behaved, Gaussian light curve that we retained in our final sample.  The bottom panels show an observation toward the X-ray binary 4U1822-37 (Obs. ID 401051010), but excluding the point source from the extraction region.  We excluded this observation from the sample because the light curve shows clear episodic variations due to residual X-ray binary emission.
}
\label{figure.lc_regfilter_chap_fb}
\end{center}
\end{figure*}



\begin{figure*}
\begin{center}
\includegraphics[width = 1.0\textwidth, keepaspectratio=true]{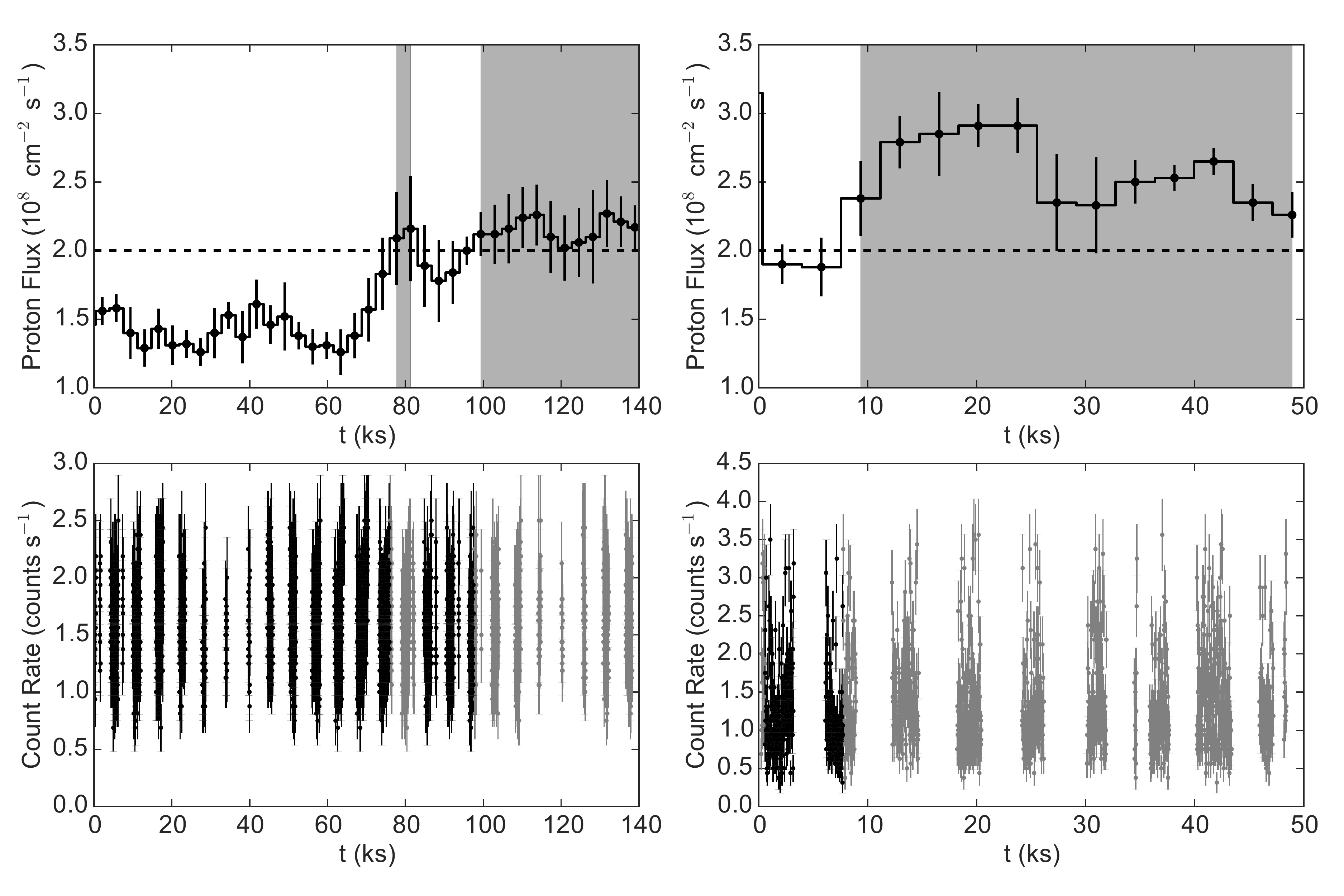}
\caption[Solar wind proton flux and X-ray light curves]{Example solar wind proton flux light curves (top panels) and 0.4--10.0 keV light curves (bottom panels) light curves for two observations in our initial sample.  Observing periods when the solar wind proton flux exceeded $2 \times 10^8$ cm$^{-2}$ s$^{-1}$ (black dashed lines in the {top} panels) were excluded.  We represent these periods with a gray background in the {top} panels and gray points in the {bottom} panels.  The {left} panels show the same Galactic bulge observation from Figure~\ref{figure.lc_regfilter_chap_fb} (Obs. ID 100011010), which had some observing time removed, but was retained in the sample because it had $\geq$5 ks of good observing time.  The {right} panels show an observation from the analysis of \citet{kataoka_etal13} (FERMI\_BUBBLE\_N2; Obs. ID 507002010).  We rejected this observation from the sample since most of it occurred during a period of high solar wind proton flux.  
}
\label{figure.lc_fluxfilter_chap_fb}
\end{center}
\end{figure*}


We compiled an initial \textit{Suzaku} target list of all observations that were publicly available as of 2015 January and near the Fermi bubbles.  This included any observations with Galactic coordinates $|l| \leq\ 25\arcdeg$ and $10\arcdeg \leq |b| \leq\ 55\arcdeg$.  There were 143 observations in this region of the sky, which we inspected for usable spectra.

Each observation includes data from the three active X-ray Imaging Spectrometer (XIS) detectors on board \textit{Suzaku} \citep{koyama_etal07}.  The detectors each have an $18\arcmin \times 18\arcmin$ field of view and a point spread function of $\approx 2 \arcmin$.  We only considered data from the back-illuminated XIS1 detector since it has a larger collecting area below 1 keV than the other detectors.  

We processed all XIS1 data using HEAsoft version 6.17 and calibration database (CALDB) files from 2015 January.  We followed the standard data reduction procedure described in \textit{The Suzaku Data Reduction Guide}.\footnote{\url{https://heasarc.gsfc.nasa.gov/docs/suzaku/analysis/abc/}}  This includes recalibrating the raw data files, screening for flickering or bad pixels, energy-scale reprocessing, and building good time interval (GTI) files.  Fortunately, the FTOOL script \texttt{aepipeline} performs all these tasks for standard \textit{Suzaku} observations.  We used \texttt{aepipeline} version 1.1.0 to generate reprocessed images, light curves, and spectra for our analysis.  

We extracted all data products using \texttt{xselect} version 2.4.  Each data set combined data from $3 \times 3$ and $5 \times 5$ editing modes where applicable.  We did not impose any non-standard criteria on the data extraction steps with the exception of the cutoff rigidity (COR) of the Earth's magnetic field.  This parameter varies throughout \textit{Suzaku's} low Earth orbit, and larger COR constraints result in fewer particle background counts.  A higher cutoff than the standard COR $>$ 4 GV has been used in previous SXRB spectral fits, and typically results in higher signal-to-noise ratio between 0.5 and 2.0 keV \citep{smith_etal07, kataoka_etal15}.  Here, we follow the suggested value from \citet{smith_etal07} and use a constraint of COR $>$ 8 GV.

Since the observational goal of this study is to measure \ovii and \oviii intensities from SXRB spectra, we removed point sources from each observation before we extracted spectra.  We inspected each XIS1 image for point sources to remove with self-defined region files.  Observations of exceptionally bright sources (i.e., where there was clear emission extending over more than $\approx5\arcmin$), extended sources (galaxy clusters, star clusters, etc.), or other anomalous features were rejected from the data set.  We also rejected any observations not taken in the standard observing window mode due to the reduction in field-of-view collecting area.  For any remaining visible point sources, we defined circular exclusion regions between 1$\arcmin$ and 4$\arcmin$ in radius centered on each source.  We then re-extracted the data products for each observation with the point source region excluded.  

The next step in our data cleaning process was light curve inspection and filtering.  We extracted 0.4--10.0 keV light curves and constructed count-rate histograms for each observation.  Our default screening criterion was to remove observing periods that were $> 2 \sigma$ from the mean count rate.  This led to a small reduction in observing time since most count-rate histograms followed an approximately Gaussian distribution.  We flagged observations that did not have an approximately Gaussian count-rate distribution, which is indicative of additional soft proton flares or residual point-source emission that may have been variable throughout the observation.  Example light curves with the various filtering tasks can be seen in Figure~\ref{figure.lc_regfilter_chap_fb}.  We also expand on this analysis step in Section~\ref{subsection.swcx_chap_fb} where we exclude observing periods with high solar wind proton flux measurements.  This light curve filtering created new GTI files, which we used to compile our initial data products.  

The procedures for point-source exclusion and light curve filtering procedures outlined above were the primary stages in our initial sample catalog.  To summarize the main screening criteria, we excluded observations that showed anomalous features in either their images or light curves or that had exceptionally bright point sources, and we filtered the images for removable point sources.  After these screening criteria, we kept observations with $\geq$5 ks of good observing time.  This resulted in 112 observations out of the original 143.

\subsection{SWCX Filtering}
\label{subsection.swcx_chap_fb}






SWCX emission can occur in any X-ray observation, but it is difficult to predict or attribute the amount of SWCX emission in individual SXRB observations \citep[e.g., ][]{carter_sembay08, carter_etal11, galeazzi_etal14, hs15}.  The emission can occur either at the interface between the solar system and interstellar medium (ISM) (heliospheric emission) or as solar wind ions pass near the Earth's neutral atmosphere (geocoronal emission).  Heliospheric emission is thought to vary with the overall solar cycle, the observed direction relative to the Sun's orbit and ecliptic plane, and the hydrogen and helium density in the neutral ISM \citep{cravens_etal01, robertson_cravens03_a, koutroumpa_etal06, koutroumpa_etal07, koutroumpa_etal11, galeazzi_etal14, hs15}.  Geocoronal emission is thought to depend on the solar wind proton flux and the observed direction relative to the magnetosheath \citep{snowden_etal04, wargelin_etal04, fujimoto_etal07, carter_sembay08, carter_etal10, carter_etal11, ezoe_etal10, ezoe_etal11, ishikawa_etal13, hs15}.  It is still unclear what the typical amount of SWCX emission is in a given X-ray observation, and models predict a wide range of \ovii and \oviii intensities depending on the parameters listed above \citep{robertson_cravens03_b,koutroumpa_etal06, koutroumpa_etal07, koutroumpa_etal11,robertson_etal06}.  For the purpose of this project, SWCX emission is considered to be contamination, and our goal is to reduce the amount of potential emission as much as possible.

Following the work of \citetalias{hs12}, we filter the observations for periods of high solar wind proton flux in an effort to reduce geocoronal SWCX.  For each observation, we gathered solar wind data from the OMNIWeb database,\footnote{\url{http://omniweb.gsfc.nasa.gov/}} which includes data from the \textit{Advanced Composition Explorer} and \textit{Wind} satellites.  The database includes solar wind densities and velocities, which we convert to fluxes.  We cross-correlated each solar wind proton flux light curve with the X-ray light curves.  Periods with solar wind flux values $> 2 \times 10^8 $ cm$^{-2}$ s$^{-1}$ were flagged and considered to potentially include SWCX emission.  We illustrate how this screening works for several example spectra in Figure~\ref{figure.lc_fluxfilter_chap_fb}.  We made new GTI files incorporating these filtering periods and the $> 2 \sigma$ count rate periods discussed above.  These GTI files were used for the final spectral extraction used in the fitting analysis.  

We point out that this filtering procedure is designed to reduce geocoronal SWCX emission, not heliospheric SWCX emission.  The models suggesting that heliospheric SWCX varies with ecliptic latitude imply that applying an ecliptic latitude cut to an observing sample may help reduce that emission.  Indeed, \citet{hs13} discuss this effect and employ this screening criterion for their study on fitting SXRB spectra.  However, the analysis in \citetalias{miller_bregman15} argues that there does not appear to be a significant enhancement of \ovii or \oviii line emission within 10$\arcdeg$ of the ecliptic plane, part of which passes through the Fermi bubbles.  Therefore, heliospheric SWCX is likely present at softer X-ray energies, but it does not appear to be a significant emission source for the oxygen lines of interest.  

This additional screening procedure can only reduce the good exposure time in a given observation.  Some observations occurred entirely during a period of high solar wind proton flux, in which case the observation was removed from the sample.  Other observations occurred entirely during a period of low solar wind proton flux, in which case the observation was unaffected.  The rest of the observations were partially contaminated, leading to a reduction of observing time.  We enforced the same good exposure time requirement noted above of $>$5 ks to keep observations in the sample.  After the default screening outlined in Section~\ref{subsection.suz_chap_fb} and this additional SWCX filtering, our sample includes 58 of the original 143 observations.  


\begin{figure}
\begin{center}
\includegraphics[width = .5\textwidth, keepaspectratio=true]{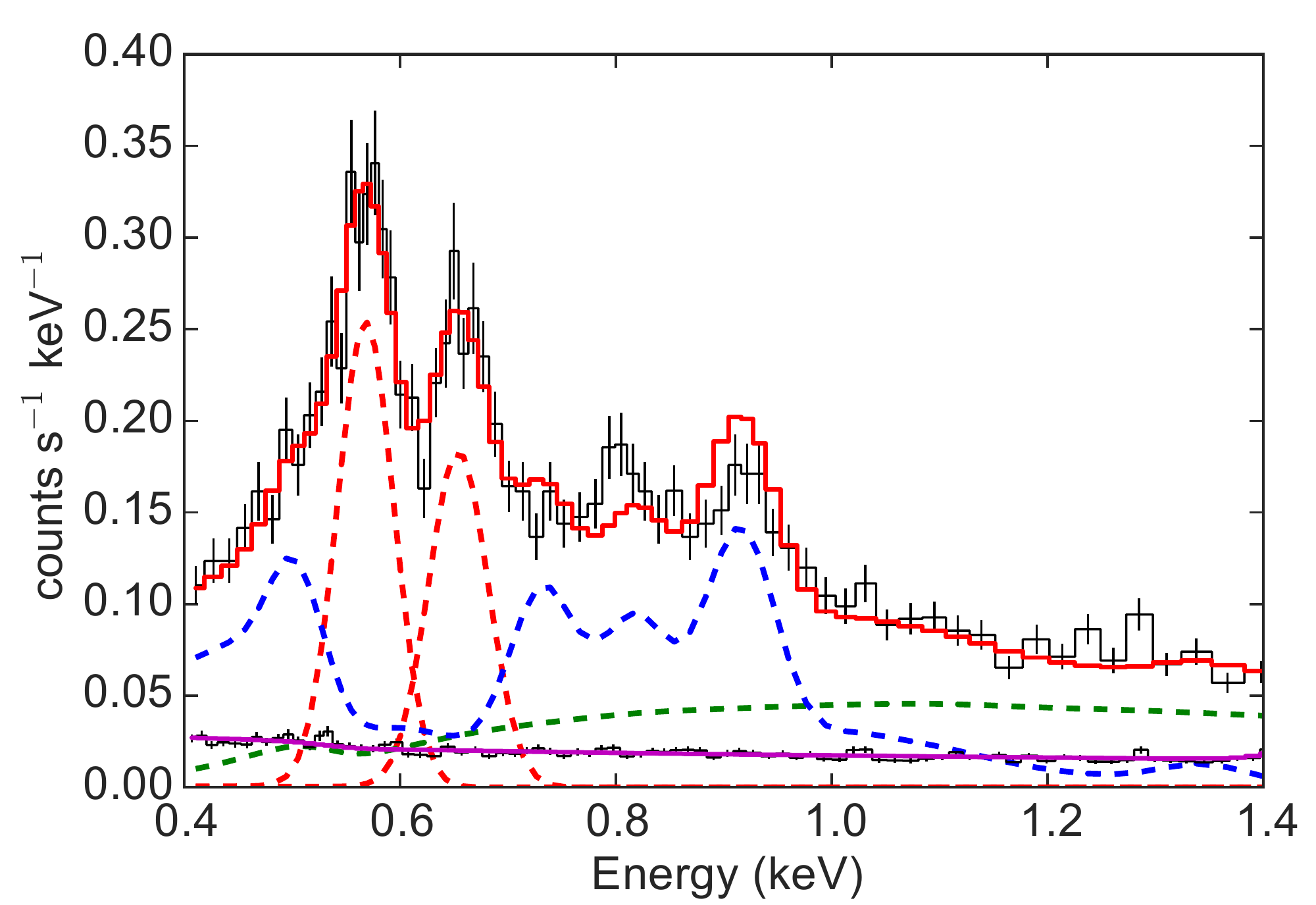}
\caption[Example SXRB spectrum for Observation ID 702028010]{Binned SXRB spectrum showing our spectral model components (Obs. ID 702028010).  The solid magenta line represents our NXB model, the green dashed line is the absorbed CXB power law, the blue dashed line is the absorbed hot gas continuum without the oxygen lines, the red dashed lines show the \ovii and \oviii lines as Gaussian components, and the solid red line represents the total model spectrum.  The oxygen lines dominate the spectrum between 0.5 and 0.7 keV.
}
\label{figure.spectrum_chap_fb}
\end{center}
\end{figure}




\begin{figure*}[t]
\begin{center}
\includegraphics[width = 1.0\textwidth, keepaspectratio=false]{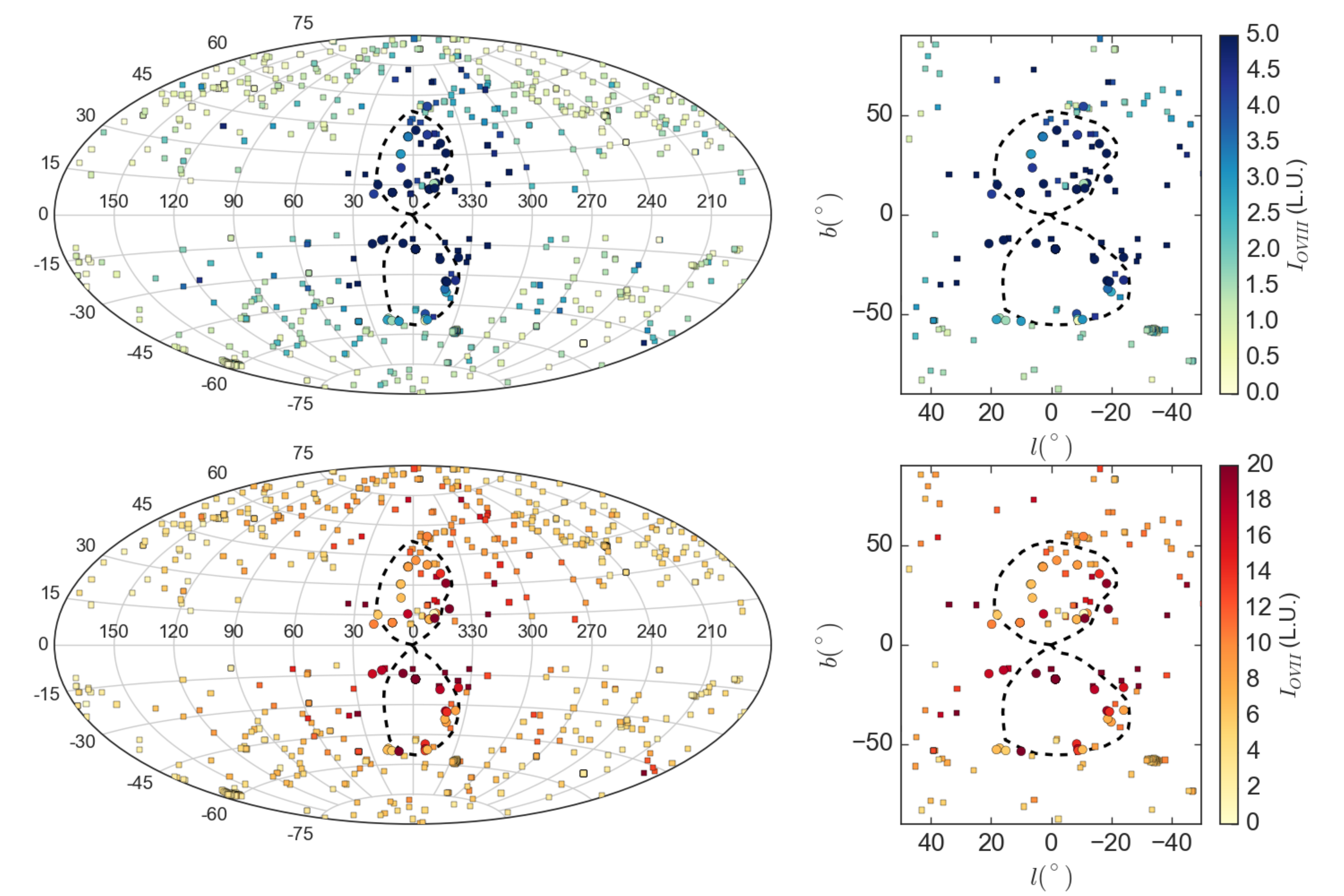}
\caption[Line intensity maps for \textit{XMM-Newton} and \textit{Suzaku} data]{All-sky Aitoff projections ({left} panels) and a projection near the Fermi bubbles ({right} panels) of our \oviii and \ovii emission line samples ({top} and {bottom} panels respectively).  The squares represent measurements from \textit{XMM-Newton} \citepalias{hs12}, the circles represent our new \textit{Suzaku} measurements, and the dashed lines represent the Fermi bubbles' gamma-ray edge.  We use the \oviii data in our model fitting process.
}
\label{figure.maps_chap_fb}
\end{center}
\end{figure*}


\subsection{Spectral Modeling}
\label{subsection.spec_mod_chap_fb}

This section outlines our spectral fitting procedure, including the response files used, our spectral model, and resultant data products.  We extracted spectra in the 0.4--5.0 keV band, which is broad enough to model the known SXRB emission sources.  Each observation had its own particle spectrum, or non X-ray background (NXB) spectrum, and response files.  We used \texttt{Xspec} version 12.9.0 for all spectral fitting, where we used the Cash statistic as our fit statistic \citep{cash79}.  Figure~\ref{figure.spectrum_chap_fb} shows an example observed spectrum and best-fit multi-component model.  Our final result includes \ovii and \oviii line intensities for each observation.  

We generated response matrices and auxiliary response files using standard \textit{Suzaku} scripts.  The script \texttt{xisrmfgen} was used to generate response matrices (RMF files) for each observation.  We used the ray-tracing program \texttt{xissimarfgen} to generate ancillary response files (ARFs) assuming a uniformly emitting source and a 20$\arcmin$ radius circle for a simulated source region.  The size of the source region acts as a normalization in our spectral fit values.

Each observation has an NXB spectrum collected by \textit{Suzaku} observations of the Earth at night.  We generated NXB spectra using the script \texttt{xisnxbgen} \citep{tawa_etal08}.  This creates an NXB spectrum using a weighted sum of \textit{Suzaku} of the Earth at night based on the exposure times.  The only unique parameter we supplied to the script is the same constraint of COR $>$8 GV that we applied to the initial data extraction.  Different SXRB studies have multiple treatments for these NXB spectra, with groups either subtracting the NXB counts from the observed spectrum \citepalias[e.g., ][]{kataoka_etal15} or including the NXB spectrum as additional data and simultaneously modeling its contribution to the full spectrum \citep[e.g., ][]{smith_etal07}.  We follow the latter methodology, meaning we fit both our observed spectrum and the NXB spectrum as one process.  

Our spectral model includes the following components: NXB spectrum, an absorbed cosmic X-ray background (CXB) or extragalactic power law, an absorbed hot gas continuum component with no oxygen emission lines, and the \ovii and \oviii lines of interest.  The absorbed components were attenuated using the \texttt{phabs} model in Xspec \citep{bc_mccammon92, yan_etal98} and had column densities fixed to values from the LAB survey \citep{kalberla_etal05}.  We also assume metal abundances from \citet{asplund_etal09} unless otherwise stated.  The rest of this section details the spectral model assumed for each source.  

The NXB model includes a contribution from particles hitting the XIS detectors that are not focused by the telescope and three instrumental lines.  For the particle spectrum, we include a power law where both the normalization and slope can vary.  The instrumental lines include an Al K line at 1.49 keV, an Si K line at 1.74 keV, and an Au M line at 2.12 keV.\footnote{\url{https://heasarc.gsfc.nasa.gov/docs/suzaku/prop_tools/suzaku_td/}}  We model each of these lines as Gaussians with widths and normalizations left to vary as free parameters and the centroids fixed.  This model component is only folded through the RMF response (as opposed to both the RMF and ARF), and it contributes to both the observed and NXB spectra.  

The CXB spectrum is typically modeled as an absorbed power law or multiple broken power laws, and is thought to be due to unresolved active galactic nuclei (AGNs).  The differences between these spectral shapes have minimal effects on the measured oxygen line intensities, since the CXB contributes $\lesssim10\%$ to the total SXRB flux below $\approx$1 keV.  Therefore, we adopt the spectral shape used by \citetalias{hs12}, which is a power law with fixed spectral index of 1.46 \citep{chen_etal97}.  These authors discuss the uncertainty in the CXB power law normalization, but argue for a nominal value of 7.9 photons cm$^{-2}$ s$^{-1}$ sr$^{-1}$ keV$^{-1}$ at 1 keV after accounting for CXB sources with $F_X < 5 \times 10^{-14}$ erg cm$^{-2}$ s$^{-1}$ in the 0.5--2.0 keV band \citep{moretti_etal03}.  We allow the CXB normalization to be a free parameter in our spectral model, but place $\pm$30\% hard boundaries on the parameter to allow for field-to-field variation.  

Although the oxygen lines are the measurement of interest, we still account for hot gas continuum emission in our model.  We model this component as an absorbed thermal APEC plasma \citep{smith_etal01} with fixed solar metallicty and without the oxygen lines.  We achieve the latter by setting the oxygen line emissivities to zero in the standard Xspec APEC files \citep{lei_etal09, hs12}.  The normalization and temperature were left as free parameters in the model.  We expected the fitted plasma temperatures to be between 0.1 and 0.3 keV, but these temperatures are typically most sensitive to the oxygen lines that we disabled.  Therefore, we let the plasma temperature vary outside this range, but with hard boundaries between 0.05 and 5 keV.

Our final spectral model components are the \ovii and \oviii emission lines.  We model each of these components as Gaussian features with the widths fixed to the instrumental resolution.  We fixed the \oviii centroid to its laboratory value of 0.6536 keV, but let the \ovii centroid vary since the line is an unresolved blend of the resonance, forbidden, and intercombination lines.  The Gaussian normalizations were also free parameters, because these represent the line strength measurements.  We point out that this spectral fitting method measures the \textit{total} emission line strengths from all emission sources (residual SWCX, LB, absorbed hot gas halo, or Fermi bubble) for each sight line.  This is why our model line intensities in Section~\ref{subsection.intensity_chap_fb} include all emission sources along each sight line.

\subsection{The Exclusion of the North Polar Spur Region}
\label{subsection.nps}

The North Polar Spur (NPS) is an extended region of enhanced X-ray emission that is cospatial with part of a larger region of enhanced radio continuum emission known as Loop I \citep{berkhuijsen_etal71,borken_iwan77,snowden_etal97}.  The X-ray enhancement is strongest at lower Galactic latitudes near $l, b \approx$ 25\arcdeg, 25\arcdeg, and gradually decreases in intensity toward the Galactic pole.  Spectral observations with \textit{XMM-Newton} and \textit{Suzaku} indicate that the plasma is hotter than the surrounding medium with $kT \approx 0.3$ keV, is depleted in C, O, Mg, Ne, Fe, and enhanced in N  \citep{willingale_etal03,miller_etal08,ursino_etal16}.  These measurements alone do not constrain the NPS distance, which makes it unclear whether the enhancement is associated with the Fermi bubbles.

Several independent methods place constraints on the NPS distance, but there is tension between the results.  \citet{sofue15} analyzed how the NPS X-ray emission varies with extinction near the Aquila Rift, and concluded that the NPS must be behind the rift and $>$1 kpc from the Sun.  \citet{puspitarini_etal14} compared three-dimensional models of the Milky Way's ISM and found no evidence for a low-density, hot cavity capable of producing the NPS X-ray emission within $\approx$200 pc of the Sun.  Alternatively, \citet{sun_etal15} mapped the NPS in polarized radio emission and constrained the distance using maps of Faraday rotation measure and depth.  They argue that the NPS emission at $b \gtrsim$50\arcdeg\ is likely within a few hundred parsecs of the Sun, while emission at lower latitudes can be either local or distant depending on the sign of the Milky Way's large-scale magnetic field.

These distance discrepancies are problematic because the NPS could be a region of compressed circumgalactic medium (CGM) material due to the Fermi bubbles' expansion, and thus a probe of their kinematics.  This scenario was suggested long before the bubbles' discovery, and models of ``biconical hypershells'' expanding away from the Galactic center can reproduce the X-ray and radio morphologies of the NPS \citep{sofue77,sofue00,sofue_etal16}.  Alternatively, models of stellar winds and supernova remnants from the Sco--Cen OB association can also reproduce the X-ray and radio morphologies, in which case the emission occurs $\approx$200 pc away from the Sun \citep{egger_aschenbach95,wolleben07}.  In this scenario, the NPS has no affiliation from the Fermi bubbles, so analyzing its emission would lead to inaccurate measured bubble properties.

In light of these uncertainties, we choose to exclude observations that are projected through the NPS in our model fitting analysis.  This is a conservative approach that prevents us from fitting hot gas emission that may be unrelated to our emission model.  Furthermore, the NPS appears to be a unique feature in terms of its location on the sky.  There should be similarly enhanced X-ray-emitting regions in the other quadrants of the sky if the NPS were due to the Fermi bubbles, but these are not seen in the \textit{ROSAT} maps.  Thus, modeling the NPS emission could lead to biased and inaccurate results.

We exclude observations near the NPS in two square regions.  Any observations within $l$ = 20\arcdeg--25\arcdeg, $b$ = 26\arcdeg--34\arcdeg, or $l$ = 4\arcdeg--20\arcdeg, $b$ = 40\arcdeg--57\arcdeg\ were excluded.  This led to the exclusion of 27 observations from our model fitting procedure.


\begin{deluxetable*}{l r r r c r r}[t!]
\tablewidth{\textwidth}
\tablecaption{\textit{Suzaku} Data}
\tablehead{
  \colhead{Obs. ID} &
  \colhead{$l$} &
  \colhead{$b$} &
  \colhead{$t_{exp}$\tablenotemark{a}} &
  \colhead{$N_{H}$\tablenotemark{b}} &
  \colhead{$I_{OVII}$} &
  \colhead{$I_{OVIII}$} \\
  \colhead{} &
  \colhead{(deg)} &
  \colhead{(deg)} &
  \colhead{(ks)} &
  \colhead{($10^{20}$ cm$^{-2}$)} &
  \colhead{(L.U.)\tablenotemark{c}} &
  \colhead{(L.U.)\tablenotemark{c}}
  }  
\startdata

100011010 & 341.0 & 18.0 & 28.7 & 6.47 & 31.37 $\pm$ 0.84 & 20.93 $\pm$ 0.57 \\
100041020 & 358.6 & -17.2 & 20.0 & 6.57 & 17.05 $\pm$ 1.10 & 9.12 $\pm$ 0.65 \\
101009010 & 358.6 & -17.2 & 9.3 & 6.57 & 18.17 $\pm$ 1.95 & 10.54 $\pm$ 1.13 \\
102015010 & 358.6 & -17.2 & 20.1 & 6.57 & 20.35 $\pm$ 1.59 & 9.95 $\pm$ 0.87 \\
103006010 & 358.6 & -17.2 & 18.2 & 6.57 & 18.76 $\pm$ 1.79 & 9.14 $\pm$ 0.93 \\
104022010 & 358.6 & -17.2 & 5.6 & 6.57 & 22.13 $\pm$ 3.70 & 11.92 $\pm$ 1.97 \\
104022020 & 358.7 & -17.1 & 22.0 & 6.57 & 20.67 $\pm$ 1.77 & 9.91 $\pm$ 0.93 \\
104022030 & 358.6 & -17.2 & 19.7 & 6.57 & 20.78 $\pm$ 2.05 & 11.19 $\pm$ 1.03 \\
105008010 & 358.6 & -17.2 & 27.4 & 6.57 & 21.57 $\pm$ 1.76 & 12.08 $\pm$ 0.92 \\
106009010 & 358.6 & -17.2 & 23.5 & 6.57 & 37.77 $\pm$ 4.63 & 12.39 $\pm$ 1.39 \\
107007010 & 358.6 & -17.2 & 29.6 & 6.57 & 19.43 $\pm$ 1.58 & 11.54 $\pm$ 0.92 \\
107007020 & 358.6 & -17.2 & 29.0 & 6.57 & 39.46 $\pm$ 2.29 & 11.56 $\pm$ 0.85 \\
108007010 & 358.6 & -17.2 & 29.7 & 6.57 & 23.77 $\pm$ 1.64 & 11.64 $\pm$ 0.89 \\
108007020 & 358.6 & -17.2 & 27.5 & 6.57 & 22.71 $\pm$ 1.70 & 11.04 $\pm$ 0.89 \\
109008010 & 358.6 & -17.2 & 26.8 & 6.57 & 23.60 $\pm$ 1.74 & 11.86 $\pm$ 0.93 \\
401001010 & 344.0 & 35.7 & 26.9 & 7.36 & 14.15 $\pm$ 0.79 & 8.52 $\pm$ 0.48 \\
401041010 & 348.1 & 15.9 & 7.1 & 12.90 & 9.03 $\pm$ 1.60 & 8.30 $\pm$ 1.16 \\
402002010 & 5.0 & -14.3 & 25.4 & 8.72 & 23.17 $\pm$ 1.47 & 14.19 $\pm$ 0.87 \\
402038010 & 6.3 & 23.6 & 55.7 & 12.10 & 6.79 $\pm$ 0.64 & 4.28 $\pm$ 0.38 \\
403024010 & 349.2 & 15.6 & 24.3 & 13.90 & 0.82 $\pm$ 0.82 & 1.50 $\pm$ 0.56 \\
403026010 & 17.9 & 15.0 & 22.9 & 16.20 & 6.03 $\pm$ 1.27 & 7.18 $\pm$ 0.82 \\
403034020 & 351.5 & 12.8 & 7.6 & 16.70 & 10.91 $\pm$ 2.16 & 7.64 $\pm$ 1.43 \\
403034060 & 351.5 & 12.8 & 9.4 & 16.70 & 7.22 $\pm$ 1.79 & 5.29 $\pm$ 1.19 \\
405032010 & 2.6 & 15.5 & 15.5 & 12.30 & 17.23 $\pm$ 2.00 & 11.38 $\pm$ 1.07 \\
406033010 & 19.8 & 10.4 & 36.5 & 19.90 & 10.49 $\pm$ 0.95 & 4.22 $\pm$ 0.54 \\
406042010 & 15.9 & -12.7 & 6.0 & 9.18 & 17.41 $\pm$ 3.48 & 10.22 $\pm$ 2.08 \\
503082010 & 17.2 & -51.9 & 22.4 & 1.48 & 6.26 $\pm$ 0.95 & 1.50 $\pm$ 0.43 \\
503083010 & 18.2 & -52.6 & 19.5 & 1.56 & 6.86 $\pm$ 1.10 & 2.84 $\pm$ 0.49 \\
507011010 & 351.5 & -49.8 & 7.7 & 1.51 & 14.31 $\pm$ 2.52 & 4.05 $\pm$ 1.09 \\
507012010 & 351.2 & -52.3 & 6.8 & 1.03 & 17.14 $\pm$ 3.09 & 2.31 $\pm$ 1.03 \\
507013010 & 351.0 & -53.1 & 5.9 & 1.10 & 13.30 $\pm$ 3.19 & 0.43 $\pm$ 1.00 \\
701029010 & 349.6 & -52.6 & 75.4 & 1.06 & 7.01 $\pm$ 0.56 & 3.07 $\pm$ 0.31 \\
701056010 & 10.4 & 11.2 & 41.5 & 19.60 & 5.61 $\pm$ 0.68 & 5.79 $\pm$ 0.43 \\
701094010 & 351.3 & 40.1 & 84.8 & 6.90 & 9.13 $\pm$ 0.47 & 3.92 $\pm$ 0.27 \\
702028010 & 20.7 & -14.5 & 17.4 & 7.35 & 16.49 $\pm$ 1.33 & 6.49 $\pm$ 0.66 \\
702118010 & 335.9 & -21.3 & 49.6 & 6.45 & 15.82 $\pm$ 0.95 & 9.84 $\pm$ 0.56 \\
703005010 & 351.3 & 40.1 & 28.2 & 6.89 & 8.93 $\pm$ 1.08 & 4.26 $\pm$ 0.58 \\
703015010 & 335.8 & -32.8 & 24.8 & 3.16 & 8.55 $\pm$ 1.26 & 4.52 $\pm$ 0.62 \\
703030010 & 15.1 & -53.1 & 64.1 & 1.95 & 5.70 $\pm$ 0.65 & 1.59 $\pm$ 0.33 \\
704010010 & 340.1 & -38.7 & 29.3 & 6.07 & 8.64 $\pm$ 1.06 & 2.70 $\pm$ 0.53 \\
705014010 & 345.6 & -22.4 & 34.5 & 4.87 & 17.44 $\pm$ 1.32 & 10.65 $\pm$ 0.69 \\
705026010 & 358.2 & 42.5 & 7.7 & 7.39 & 9.59 $\pm$ 4.07 & 7.48 $\pm$ 1.27 \\
705028010 & 341.2 & -37.1 & 17.8 & 5.52 & 7.59 $\pm$ 1.30 & 3.57 $\pm$ 0.70 \\
705041010 & 10.4 & 11.2 & 102.7 & 19.60 & 6.49 $\pm$ 0.60 & 5.91 $\pm$ 0.36 \\
706010010 & 341.6 & 30.8 & 40.4 & 8.33 & 19.27 $\pm$ 1.27 & 7.64 $\pm$ 0.69 \\
706044010 & 348.8 & 13.3 & 6.2 & 13.60 & 21.48 $\pm$ 3.89 & 9.98 $\pm$ 1.67 \\
707035010 & 10.4 & 11.2 & 33.7 & 19.60 & 9.52 $\pm$ 1.03 & 5.19 $\pm$ 0.64 \\
707035020 & 10.4 & 11.2 & 52.9 & 19.60 & 10.31 $\pm$ 0.90 & 5.87 $\pm$ 0.50 \\
801001010 & 2.7 & 39.3 & 15.6 & 8.41 & 7.13 $\pm$ 1.22 & 2.87 $\pm$ 0.65 \\
801002010 & 2.9 & 39.3 & 12.0 & 8.34 & 6.69 $\pm$ 1.40 & 3.23 $\pm$ 0.74 \\
801003010 & 2.9 & 39.1 & 14.2 & 8.34 & 4.33 $\pm$ 1.09 & 3.39 $\pm$ 0.74 \\
801004010 & 2.7 & 39.1 & 12.6 & 8.40 & 8.81 $\pm$ 1.44 & 3.39 $\pm$ 0.77 \\
801094010 & 341.4 & -33.1 & 5.2 & 4.60 & 18.43 $\pm$ 2.41 & 5.22 $\pm$ 1.03 \\
803022010 & 6.9 & 30.5 & 22.5 & 10.90 & 0.57 $\pm$ 1.04 & 1.70 $\pm$ 0.80 \\
803071010 & 6.6 & 30.5 & 94.0 & 10.80 & 6.65 $\pm$ 0.57 & 3.00 $\pm$ 0.31 \\
805036010 & 340.6 & -33.6 & 22.5 & 4.33 & 14.19 $\pm$ 1.79 & 5.50 $\pm$ 0.95 \\
807048010 & 10.0 & -53.5 & 51.2 & 1.27 & 23.45 $\pm$ 1.80 & 2.97 $\pm$ 0.38 \\
807062010 & 349.3 & 54.4 & 5.2 & 2.90 & 10.37 $\pm$ 2.73 & 4.13 $\pm$ 1.41
\enddata
\label{table.suz_data_chap_fb}

\tablecomments{Table summarizing our \textit{Suzaku} emission line sample.  The columns represent the observation ID, the Galactic coordinates of the observation, the good XIS1 exposure time, the Galactic hydrogen column density, and the oxygen emission lines of interest.  The line intensity uncertainties are the 1$\sigma$ statistical uncertainties from \texttt{Xspec}.  }

\tablenotetext{a}{The total good XIS1 exposure time after our default light-curve filtering and additional flux-filtering to remove geocoronal SWCX emission.}
\tablenotetext{b}{We use hydrogen column densities from the LAB survey \citep{kalberla_etal05}.  These columns are used in our spectral fitting to absorb continuum emission and in our model line intensity calculation to attenuate emission from the hot gas halo and Fermi bubble/shell.}
\tablenotetext{c}{1 L.U. = 1 Line Unit = 1 photons s$^{-1}$ cm$^{-2}$ sr$^{-1}$}.

\end{deluxetable*}


\subsection{Data Summary}
\label{subsection.data_summary_chap_fb}

The final \textit{Suzaku} sample with spectral fitting results can be seen in Table~\ref{table.suz_data_chap_fb}.  We include the observation ID, the sight line in Galactic coordinates, the good XIS exposure time, and the oxygen line strengths with their 1$\sigma$ uncertainties in Line Units (L.U.).  The emission line measurements presented here are designed to have as little SWCX as possible, while containing only emission from astrophysical sources of interest (i.e., Galactic hot gas halo and Fermi bubbles).  We also outlined data reduction, extraction, and cleaning prodecures as similar as possible to the work by \citetalias{hs12}, such that this sample and the \textit{XMM-Newton} data are processed in a uniform way.

Our total data sample used in our astrophysical modeling combines the \textit{XMM-Newton} and \textit{Suzaku} measurements.  There are 683 \textit{XMM-Newton} measurements in total distributed across the sky, with 34 projected near the Fermi bubbles.  The \textit{Suzaku} data are exclusively projected near the Fermi bubbles, and there are 58 measurements included here.  Figure~\ref{figure.maps_chap_fb} shows all-sky maps of the oxygen emission line strengths.

\section{Model Overview}
\label{section.model_chap_fb}

In this section, we define our parametric astrophysical models and assumptions.  The SXRB is known to have at least two plasma sources---a ``local'' source within $\approx$300 pc from the Sun and a ``distant'' source at  $\gtrsim$5 kpc from the Sun.  These sources have all been modeled in different ways, resulting in different inferences on their underlying emission properties.  The Fermi bubbles have not been considered in most SXRB modeling studies, with the exception of recent studies by \citetalias{kataoka_etal15}.  Here, we identify all emission sources in our model, and justify our choices for the underlying source distributions.  

We point out that this work is an advancement over the modeling work presented in \citetalias{miller_bregman15}, who used the same \textit{XMM-Newton} data discussed in Section~\ref{subsection.xmm_chap_fb} to constrain the structure of the  Milky Way's hot gas halo.  The model used in that study is identical to the model outlined below, with the exception of the Fermi bubble emission source.  We summarize these models in Sections~\ref{subsection.lb_mod_chap_fb} and ~\ref{subsection.halo_mod_chap_fb}, but we refer the reader to \citetalias{miller_bregman15} for additional explanation of the model choice.


\subsection{LB / Residual SWCX Model}
\label{subsection.lb_mod_chap_fb}

The ``local'' emission source has been argued to include emission from both the LB and SWCX based on numerous shadowing experiments \citep{galeazzi_etal07,koutroumpa_etal07,koutroumpa_etal11,smith_etal07} and studies of the \textit{ROSAT} 1/4 keV band \citep{kuntz_snowden00, galeazzi_etal14}.  As discussed in Section~\ref{subsection.swcx_chap_fb}, SWCX emission is difficult to predict or quantify; however, the flux-filtering techniques are designed to reduce its contribution to our measured line strengths.  The physical properties of the LB have also been debated, with some studies arguing that the LB is volume-filled with $\sim10^6$ K gas \citep{smith_etal07} and others arguing that the emission comes primarily from the bubble edges about 100--300 pc away \citep{lallement_etal03,welsh_shelton09}.  Regardless of these differences, our goal is to choose a parameterization that characterizes the emission from this source.

We parameterize the LB as a volume-filled plasma with a constant density and temperature and a size varying between 100 and 300 pc.  This follows interpretation from \citet{smith_etal07}, who conducted SXRB modeling with \textit{Suzaku} on the nearby molecular cloud MBM12.  Under the assumption of a volume-filled plasma, these authors concluded that the LB has a temperature of $1.2 \times 10^6$ K and a density of 1--4$\times 10^{-3}$ cm$^{-3}$.  In our model, we fix the plasma temperature to this value and let the density, $n_{LB}$, be a free parameter.

While we include an Local Bubble emission source in our model for completeness, it is unlikely to have a significant impact on our results.  The ``local'' plasma source is known to contribute more to the \textit{ROSAT} 1/4 keV band than to the \textit{ROSAT} 3/4 keV band \citep[e.g., ][]{snowden_etal97,kuntz_snowden00}.  This implies that the ``local'' emission source should produce more \ovii than \oviii in a given observation.  Shadowing spectroscopic observations verify this, and show there is minimal ($\lesssim$0.5 L.U.) \oviii due to ``local'' sources \citep{koutroumpa_etal07,koutroumpa_etal11,smith_etal07}.  Furthermore, \citetalias{miller_bregman15} showed that this LB model effectively has no contribution to the \oviii emission lines from the \textit{XMM-Newton} sample discussed in Section~\ref{subsection.xmm_chap_fb}.  Our modeling work below focuses on \oviii emission lines, so we do not believe that this LB parameterization will affect our results.

\subsection{Hot Halo Model}
\label{subsection.halo_mod_chap_fb}

We assume that the Milky Way's  ``extended'' hot gas plasma structure is dominated by a spherical, volume-filling halo of material extending to the virial radius, as opposed to the alternative assumption of an exponential disk morphology with scale height between 5 and 10 kpc.  The latter structure is believed to form from supernovae in the disk \citep[e.g., ][]{norman_ikeuchi89, joung_maclow06, hill_etal12} and can reproduce X-ray absorption and emission line strengths in several individual sight lines \citep{yao_wang05,yao_wang07,yao_etal09_b,hagihara_etal10}.  However, numerous studies have shown that a spherical, extended morphology due to shock-heated gas from the Milky Way's formation reproduces a multitude of observations \citep[e.g., ][]{white_frenk91,cen_ostriker06, fukugita_peebles06}.  These include ram-pressure stripping of dwarf galaxies \citep{blitz_robishaw00,grcevich_putman09,gatto_etal13}, the pulsar dispersion measure toward the Large Magellanic Cloud \citep{anderson_bregman10,fang_etal13}, and the aggregate properties of oxygen absorption and emission lines distributed in multiple sight lines across the sky \citep{bregman_ld07,gupta_etal12,miller_bregman13,miller_bregman15,faerman_etal16}.  This distribution has been proven to reproduce most of the \oviii emission line intensities from the \textit{XMM-Newton} portion of the sample, thus justifying its use in this modeling work.

Our parameterized density distribution follows a spherical $\beta$-model, which assumes that the hot gas is approximately in hydrostatic equilibrium with the Milky Way's dark-matter potential well.  The $\beta$-model has also been used to fit X-ray surface brightness profiles around early-type galaxies \citep[e.g., ][]{osullivan_etal03} and massive late-type galaxies \citep{anderson_bregman11,dai_etal12,bogdan_etal13b,bogdan_etal13a,anderson_etal16}.  The model is defined as

\begin{equation}
\label{eq.beta_model_chap_fb}
n(r) = {n_o}({1 + ({r}/{r_c})^2})^{-{3\beta}/{2}},
\end{equation}
	
\noindent where $r$ is the galactocentric radius, $n_\circ$ is the central density, $r_c$ is the core radius ($\lesssim$5 kpc), and $\beta$ defines the slope (typically between 0.4 and 1.0).  The previous modeling by \citetalias{miller_bregman15} was limited to using an approximate form of this model in the limit where $r \gg r_c$, since they specifically did not include observations near the expected $r_c$.  This resulted in constraints on a power-law density distribution:

\begin{equation}
\label{eq.beta_model_approx_chap_fb}
n(r) \approx \frac{n_or_c^{3\beta}}{r^{3\beta}}.
\end{equation}
	
\noindent The emission line sample in this study includes 33 sight lines that pass within 20$\arcdeg$ of the Galactic center, so we present model results assuming both distributions.  The net effect of this will be for the power-law model to produce more halo emission for sight lines near the Galactic center than the usual $\beta$-model since the density continues to increase at small $r$ instead of approach $n_o$ for $r \lesssim r_c$.  

We assume the halo gas is isothermal with a temperature of log($T_{halo}$) = 6.30, or $T_{halo} = 2 \times 10^6$ K.  This temperature is characteristic of the Milky Way's virial temperature, and thus consistent with the picture in which the ``extended'' plasma is spherical and extended to $r_{vir}$.  This temperature is also constrained by observations.  \citet{hs13} provide the strongest observational constraints on the plasma temperature, as they fit SXRB spectra for 110 high-latitude ($|b| > 30\arcdeg$) sight lines from the \citetalias{hs12} sample.  They fit the spectra with thermal APEC plasma models and found a narrow range of plasma temperatures (median and interquartile range of $2.22 \pm 0.63 \times 10^6$ K).  These results suggest that the plasma is nearly isothermal, thus validating our assumption.

\subsection{Fermi Bubble Geometry}
\label{subsection.geometry_chap_fb}


\begin{figure*}[t!]
\begin{center}
\includegraphics[width = 1.0\textwidth, keepaspectratio=true]{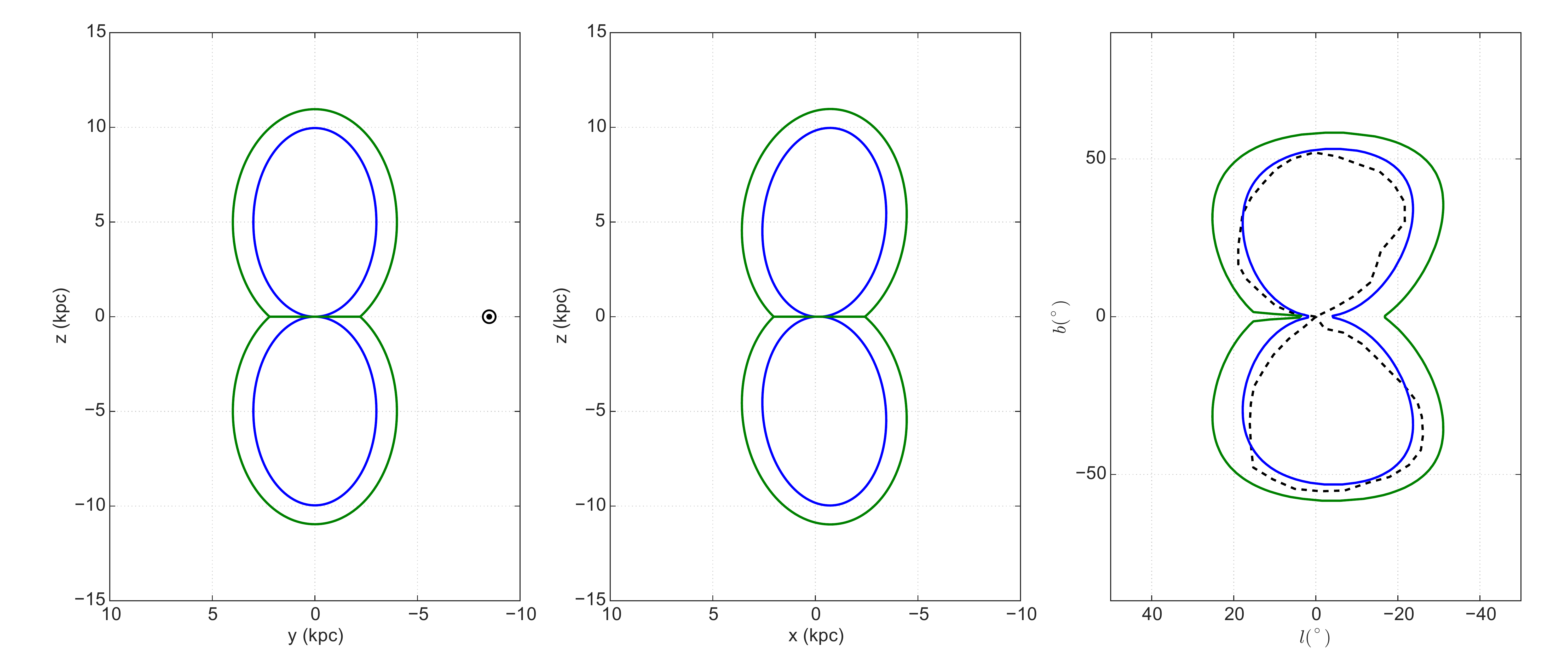}
\caption[Fermi bubble geometry in physical and projected coordinates]{Outlines of our volume-filled (blue lines) and shell (green lines) model distributions in physical galactocentric coordinates ({left} and {center}) and projected Galactic coordinates ({right}).  The {left} panel shows a side view of the structure in the $l = 0\arcdeg$ plane (the $\odot$ symbol represents the Sun), while the {center} panel shows a face-on view at the Galactic center.  The {right} panel indicates that our volume-filled distribution creates a projection consistent with the bubbles' observed gamma-ray outline (black dashed line).
}
\label{figure.geometry_chap_fb}
\end{center}
\end{figure*}


Our Fermi bubble structure includes two components: a volume-filled ellipsoid and a shell surrounding the ellipsoid.  These components are designed to parameterize the outer regions of galactic outflows, where the volume-filled structure includes hot, shocked wind material, and the shell includes shocked ISM/CGM material (see Section~\ref{subsubsection.wind_chap_fb} for an overview of galactic wind morphology).  Hard X-ray emission bounds the bubbles at low Galactic latitudes, verifying that there is a distinct shell structure of thermal gas surrounding the bubbles \citep{bh_cohen03,su_etal10}.  The flat gamma-ray intensity distribution on the sky indicates that non-thermal particles fill the bubbles in a quasi-spherical volume, and we include a thermal gas component in this region.

We define the bubble volume as a three-dimensional ellipsoid designed to match the bubbles' projected gamma-ray edge on the sky.  Each bubble (positive and negative Galactic latitudes) is centered at $|z|$ = 5 kpc away from the Galactic plane, has a semi-major axis of 5 kpc, and has both minor axes set to 3 kpc.  We also tilt each bubble $5\arcdeg$ toward negative longitudes to match the slight asymmetry observed in the bubble shape.  Figure~\ref{figure.geometry_chap_fb} shows this bubble volume in physical and projected coordinates.

The shell volume is defined in the same way as the bubble volume, but with a thickness of $\approx$1 kpc away from the bubble surface.  This implies that the shell ellipsoids are also centered at $|z|$ = 5 kpc from the Galactic plane and have semimajor axes of 6 kpc and minor axes of 4 kpc.  The region inside this surface but outside the bubble surface is considered to be the shell region.  We note that this parameterization is different from the modeling work presented by \citetalias{kataoka_etal15}, who considered Fermi bubble emission from only two angled shells (one for each bubble) with inner and outer radii of 3 and 5 kpc.  However, Figure~\ref{figure.geometry_chap_fb} indicates that our bubble volume is consistent with the projected bubble outline, and the expected galactic wind morphology suggests that there should be at least two distinct outflow regions we can observe (the shocked wind and shocked ISM/CGM).  Therefore, we feel that our choice of bubble volume is reasonable given the observational constraints available at this time.

\subsection{Fermi Bubble Density and Temperature}
\label{subsection.fb_nt_chap_fb}

We assume that the bubble and shell components each have constant electron densities, defined as $n_{FB}$ and $n_{shell}$.  This parameterization is useful since it is simple, yet still allows us to constrain the average thermal gas densities.  Simulations suggest that the bubbles have some thermal gas substructure \citep[e.g., ][]{yang_etal12}, and we did explore more sophisticated models with density gradients away from the Galactic plane or from the bubble edges.  However, the data did not provide statistically significant constraints with these profiles (any gradient parameter was consistent with a constant-density profile within the 1$\sigma$ uncertainties).  These constant-density models should be considered a valuable first step when analyzing the bubbles' thermal gas structure.  

The bubble and shell are likely hotter than the surrounding medium, so we assume that each component has a characteristic temperature $\geq 2 \times 10^6$ K.  Like the bubble and shell densities, each component has a constant temperature ($T_{FB}$ and $T_{shell}$).  However, these temperatures are each initially fixed to $3 \times 10^6$ K during the model fitting process.  The temperatures are not free parameters in our models because the calculated line intensity scales with density and temperature as $I \propto n^2 \epsilon(T)$, where $\epsilon$ has a temperature dependence.  Since we explicitly model a sample of \oviii emission line intensities, the density and temperature parameters would be degenerate with each other.  Section~\ref{section.results_chap_fb} discusses how we constrain the bubble and shell temperatures by looking at the distribution of \oviii/\ovii line ratios for different assumed temperatures.

Our modeling also implicitly assumes that the Fermi bubble and shell plasmas have solar metallicities.  This implies that our bubble density parameters are degenerate with the assumed metallicity since the plasma emission measure scales linearly with metallicity.  We make this choice because there are no direct observational constraints on the plasma metallicity inside the bubbles.  The SXRB modeling from \citet{kataoka_etal13} advocates for a sub-solar bubble metallicity of $Z \approx 0.2 Z_{\odot}$, but this value is weakly constrained due to photon statistics, and the spectral fits represent the emission measure-weighted spectrum from the bubbles and Galactic hot halo (see discussion in Section~\ref{subsubsection.softxray_chap_fb}).  Thus, it is difficult to interpret whether the bubbles' plasma is enriched or sub-solar.  The bubbles' abundance ratios can be diagnostics for how they formed \citep{inoue_etal15}, but a detailed abundance analysis is beyond the scope of this work.

\subsection{Line Intensity Calculation}
\label{subsection.intensity_chap_fb}


\begin{figure*}[t]
\begin{center}
\includegraphics[width = 1.\textwidth, keepaspectratio=true]{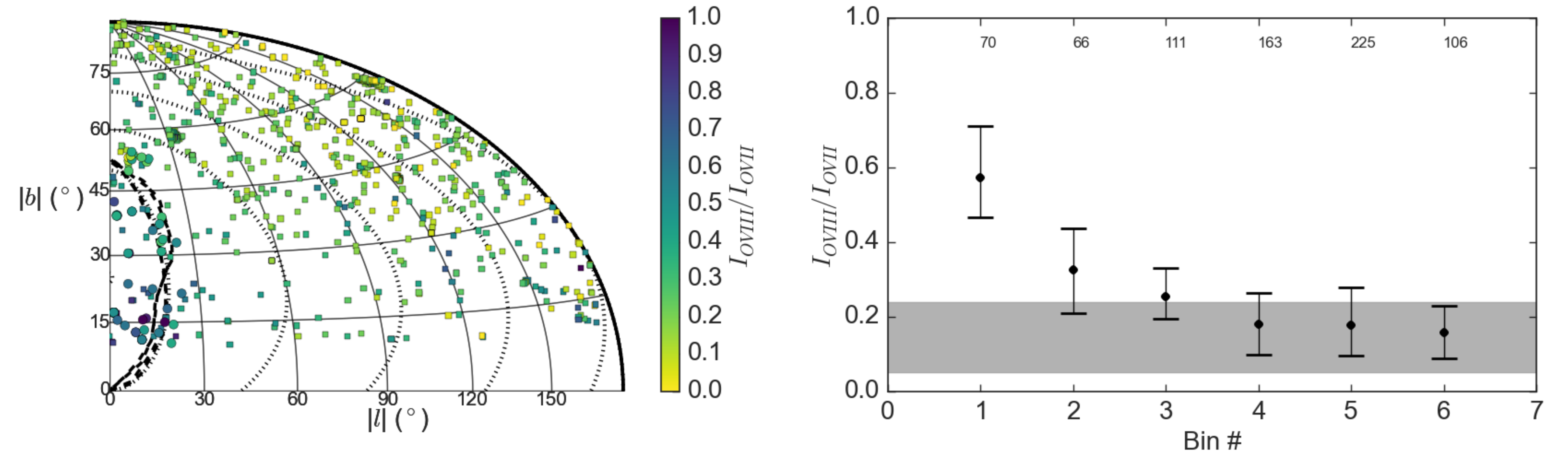}
\caption[\oviii/\ovii ratios binned on the sky]{
{Left:} measured \oviii/\ovii ratios folded into one quadrant of an Aitoff projection with the circles and squares representing \textit{Suzaku} and \textit{XMM-Newton} observations respectively.  The black dashed lines represent the observed Fermi bubble outline while the black dotted lines represent edges used to bin the data.  The line ratios are systematically larger for sight lines passing near the Fermi bubbles compared to sight lines near the Galactic pole or anti-center.  
{Right:} median and interquartile ranges of line ratios binned on the sky.  The bin edges are the black dotted lines in the {left} panel, with the first bin including all observations within the Fermi bubbles.  The top of the gray band represents the line ratio expected for a plasma with $T = 2 \times 10^6$ K, and the bottom includes a contribution from a cooler LB plasma source or SWCX.  The observations in the first bin have significantly larger line ratios than the expected range in the gray band, indicating the presence of a plasma at $> 2 \times 10^6$ K.
}
\label{figure.map_quarterbin_chap_fb}
\end{center}
\end{figure*}


Calculating model line intensities depends on the density and temperature profile along each line of sight.  For any given Galactic coordinate ($l,b$), we divide the line of sight into cells extending to the virial radius ($r_{vir}$ = 250 kpc).  Each cell position along the line of sight ($s$) is converted to Galactic coordinates ($R$, $z$, $r$) by the standard equations:

\begin{equation}
R^2 = R_{\odot}^2 + s^2\cos(b)^2 - 2sR_{\odot}\cos(b)\cos(l)
\end{equation}
\vspace{-.3cm}
\begin{equation}
z^2 = s^2\sin(b)^2
\end{equation}
\vspace{-.3cm}
\begin{equation}
r^2 = R^2 + z^2,
\end{equation}
	
\noindent where $R_{\odot}$ = 8.5 kpc is the Sun's distance from the Galactic center.  We assign a density and temperature to each cell based on its set of Galactic coordinates and the assumed model parameters.  The hot halo profile described in Section~\ref{subsection.halo_mod_chap_fb} sets the density and temperature for cells outside the shell volume.  The parameters $n_{shell}$ and $T_{shell}$ set the density and temperature for cells within the shell volume, while $n_{FB}$ and $T_{FB}$ set the density and temperature for cells within the bubble volume.  Therefore, sight lines not passing through the bubbles include only halo emission, while sight lines passing through the bubbles include emission from the hot gas halo, bubble, and shell.  

We assume an optically thin plasma in collisional ionization equilibrium to calculate all line intensities.  Given a line-of-sight density and temperature profile, the model line intensity is defined as

\begin{equation}
\label{eq.intensity_thin_chap_fb}
I (l,b) = \frac{1}{4 \pi} \int n_e(s)^2 \epsilon(T(s)) ds,
\end{equation}

\noindent where $n_e(s)$ is the line-of-sight electron density, $T(s)$ is the line-of-sight temperature profile, and $\epsilon(T)$ is the volumetric line emissivity for a thermal APEC plasma.  We use AtomDB version 2.0.2 for all line emissivities \citep{foster_etal12}, and characteristic values for \oviii (in photons cm$^3$ s$^{-1}$) are $\epsilon(T_{halo}) = 1.45 \times 10^{-15}$ and $\epsilon(3 \times 10^6\ \text{K}) = 3.84 \times 10^{-15}$.  Although believed to be minimal, the intensity contribution from the LB is defined as

\begin{equation}
I_{LB} (l,b) = \frac{n_{LB}^2 L(l,b) \epsilon(T_{LB})}{4 \pi},
\end{equation}

\noindent where $L(l,b)$ defines the LB path length ($\approx$100--300 pc; see \citet{lallement_etal03} and \citetalias{miller_bregman15}).  The total line intensity is thus defined as

\begin{equation}
I_{total} (l,b) = I_{LB} + e^{-\sigma N_{HI}}(I_{halo} + I_{FB} +I_{shell}),
\end{equation}

\noindent where the exponential term accounts for attenuation due to neutral hydrogen in the disk, $N_{HI}$ is the same neutral hydrogen column assumed for each sight line in in the spectral fitting procedure, and $\sigma$ is the \ion{H}{1} absorption cross section \citep{bc_mccammon92, yan_etal98}.  Thus, our model line intensities are comparable to the total observed line intensities.


\begin{figure*}
\begin{center}
\includegraphics[width = 1.\textwidth]{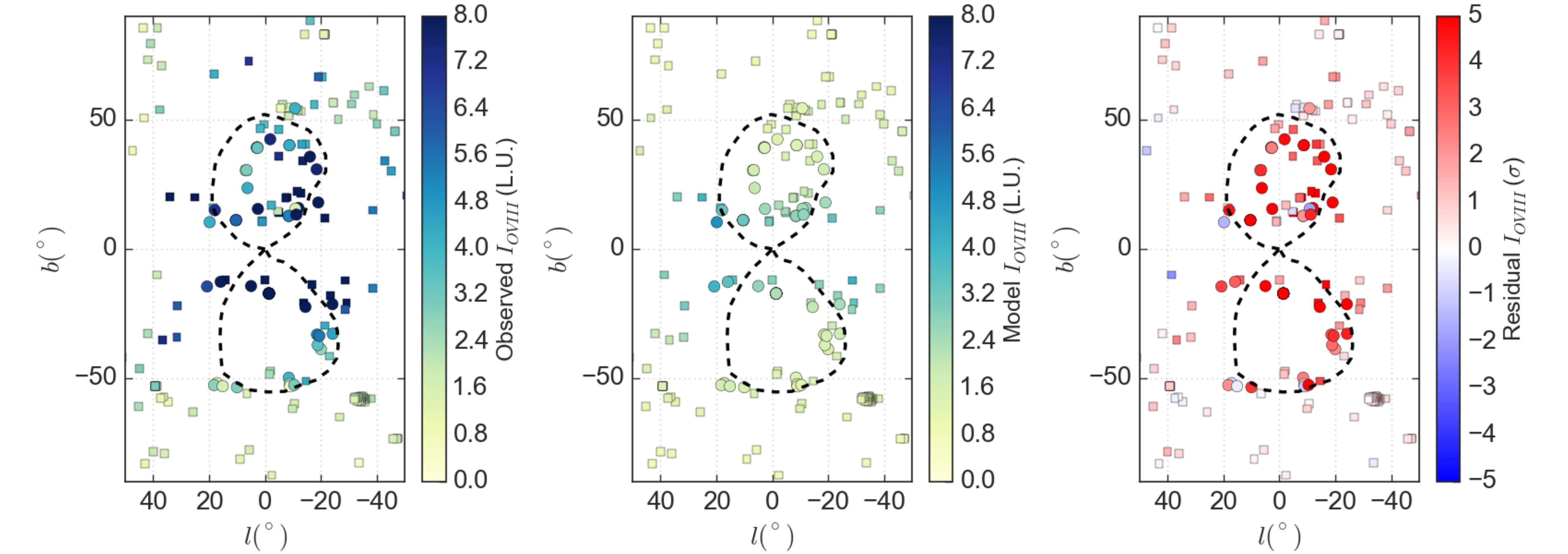}
\caption[\oviii residual maps near the Fermi bubbles]{Observed emission line intensities ({left} panel), model emission line intensities ({center} panel), and residuals ({right} panel) for a model without a bubble or shell emission component.  The hot gas halo dominates the model \oviii emission in this case.  Sight lines passing through the bubbles have significant ($\gtrsim 3 \sigma$) positive residual \oviii emission, which we attribute to the bubbles and their interaction with the ambient hot halo medium.
}
\label{figure.residuals_chap_fb}
\end{center}
\end{figure*}


\section{Results}
\label{section.results_chap_fb}

Our results include a discussion of the \ovii and \oviii line intensity distributions along with a parametric modeling analysis.  Section~\ref{subsection.ratios_chap_fb} presents the line strength and ratio distributions on the sky for the combined \textit{XMM-Newton} and \textit{Suzaku} sample.  The latter provides model-independent evidence that the bubbles contain gas at higher temperatures than the surrounding medium ($> 2 \times 10^6 $ K).  Section~\ref{subsection.obs_model_chap_fb} builds on this evidence and the modeling work from \citetalias{miller_bregman15} to constrain the characteristic thermal gas densities and temperatures associated with the bubbles.

\subsection{Emission Line Ratios}
\label{subsection.ratios_chap_fb}

The observed \oviii/\ovii ratios in our sample can be used as crude temperature diagnostics. If the observed emission lines come from a single, cospatial plasma source, Equation~\ref{eq.intensity_thin_chap_fb} indicates that the \oviii/\ovii ratio is a direct temperature diagnostic because $I_{OVIII}/I_{OVII} \propto n^2 \epsilon_{OVIII}(T)/n^2 \epsilon_{OVII}(T) = \epsilon_{OVIII}(T)/ \epsilon_{OVII}(T)$.  The observations from SXRB spectra are more complicated since we know that multiple plasma sources exist along each line of sight.  This implies that the total observed \oviii/\ovii line ratio probes the emission measure-weighted temperature due to the various plasma sources.  However, we discussed in Section~\ref{section.model_chap_fb} how the LB is believed to produce little \oviii emission with a variable amount of \ovii emission, and the hot halo plasma is believed to be nearly isothermal at $\approx 2 \times 10^6$ K.  The expected \oviii/\ovii line ratio for a thermal plasma at this temperature is $\approx 0.25$, so the observed line ratios in our sample would be $\lesssim 0.25$ if they included only emission from the LB and hot halo.  

We explore this idea by examining the \oviii/\ovii distribution on the sky from our total observation sample.  Figure~\ref{figure.map_quarterbin_chap_fb} shows our line intensity ratios on the sky.  Inspecting the sky projection alone suggests that the line intensity ratios are systematically higher for sight lines that pass through or near the Fermi bubbles ($\approx$0.5) than for those farther away from the Galactic center ($\approx$0.2).  To quantify this, we bin the sight lines on the sky and calculate the median and interquartile range for the line ratios in each bin.  The bin edges are defined as ellipses in $l,b$ space, where the first bin includes sight lines that pass though the Fermi bubbles and subsequent bins include sight lines extending farther into the halo (see dotted lines in Figure~\ref{figure.map_quarterbin_chap_fb}).  Figure~\ref{figure.map_quarterbin_chap_fb} shows the line ratio median and interquartile range for observations in each bin.  These results clearly show that the line ratios are systematically higher for sight lines in the first bin, and are also higher than the characteristic ratio expected if the observations included just LB and hot gas halo emission (gray shaded band in Figure~\ref{figure.map_quarterbin_chap_fb}).  

These systematically larger line ratios near the Galactic center indicate the presence of hotter gas than the ambient $2 \times 10^6$ K plasma.  This interpretation is model-independent and builds upon the fact that we know the Fermi bubbles occupy a significant volume above and below the Galactic center.  While this is a useful result that relies only on observations, the observed line ratios do not encode the bubbles' detailed temperature structure due to additional emission from the LB and hot gas halo.  Nevertheless, this result motivates the modeling work below and validates the assumption that the bubbles contain gas hotter than $2 \times 10^6$ K.  

\subsection{Comparing Models with Data}
\label{subsection.obs_model_chap_fb}

As a preliminary test, we explore an emission model including only contributions from the LB and hot gas halo.  This model assumes that the bubble and shell volumes contribute no line emission, or equivalently that $n_{FB} = n_{shell} = 0$.  For the other emission components, we assume a parametric model distribution from \citetalias{miller_bregman15}.  This includes an LB density of $n_{LB} = 4 \times 10^{-3}$ cm$^{-3}$ and a hot gas density profile described by Equation~\ref{eq.beta_model_approx_chap_fb} with $n_or_c^{3\beta}$ = $1.35 \times 10^{-2}$ cm$^{-3}$ kpc$^{3\beta}$ and $\beta$ = 0.5.  This model likely \textit{overestimates} any halo emission since it assumes a power law all the way to the Galactic center, as opposed to having a flat core density.  We calculate model \oviii emission line intensities for this limiting case and compute the residual emission defined as $(I_{observed} - I_{model})/I_{error}$.  Figure~\ref{figure.residuals_chap_fb} shows how the residual emission varies on the sky, with a particular emphasis on the strong ($\gtrsim 3 \sigma$) positive residuals near the Fermi bubbles.  We interpret these residuals as missing emission due to the Fermi bubbles, which motivates the modeling procedure outlined below.  

The goal of our modeling procedure is to find the density model that is most consistent with our observed data set, including contributions from the Fermi bubble and shell components.  We quantify this consistency with the model $\chi^2$ or likelihood ($\mathcal{L} \propto$ exp$(-\chi^2 / 2)$).  We use the publicly available Markov chain Monte-Carlo (MCMC) Python package \texttt{emcee} \citep{foreman_mackey_etal13} to explore our model parameter space and find the parameters that minimize the model $\chi^2$, or maximize the model ln($\mathcal{L}$).  The output chains for each model parameter are treated as marginalized posterior probability distributions.  We define ``best-fit'' parameters as the median values for each binned chain distribution, which yields identical results to the Gaussian-fitting procedure outlined by \citetalias{miller_bregman15}, assuming the distributions are approximately Gaussian.  Thus, these best-fit parameters maximize the model likelihood, given the data.  

We considered several different model parameterizations in our model fitting process.  These included hot gas halo density models described by either a power law (Equation~\ref{eq.beta_model_approx_chap_fb}) with two free parameters (the normalization and $\beta$) or a full $\beta$-model (Equation~\ref{eq.beta_model_chap_fb}) with $r_c$ and $\beta$ left to vary.  We did not let $n_o$ vary independently in this model since the previous modeling work from \citetalias{miller_bregman15} effectively constrained the halo normalization parameter $n_or_c^{3\beta}$.  Our modeling procedure keeps this quantity fixed to $n_or_c^{3\beta} = 1.35 \times 10^{-2}$ cm$^{-3}$ kpc$^{3\beta}$, while letting the core radius vary as the free parameter.  We also experimented with fixing the hot gas halo profile with the fit values from \citetalias{miller_bregman15} or with $r_c$ = 3 kpc, but we found that this made little difference in the best-fit parameters for either the halo density profile or the bubble/shell densities.


\begin{deluxetable*}{c c c c c c c c}
\tablewidth{\textwidth}
\tablecaption{MCMC Fitting Results}
\tablehead{
  \colhead{$n_{LB}$ \tablenotemark{a}} &
  \colhead{$n_or_c^{3\beta}$} &
  \colhead{$n_o$} &
  \colhead{$r_c$} &
  \colhead{$\beta$} &
  \colhead{$n_{FB}$ \tablenotemark{b}} &
  \colhead{$n_{shell}$ \tablenotemark{b}} &
  \colhead{$\chi^{2}$ (dof)}  \\
  \colhead{($10^{-3}$ cm$^{-3}$)}                  &
  \colhead{($10^{-2}$ cm$^{-3}$ kpc$^{3\beta}$)}   &
  \colhead{($10^{-3}$ cm$^{-3}$)} &
  \colhead{(kpc)} &
  \colhead{}                                     &
  \colhead{($10^{-4}$ cm$^{-3}$)} &
  \colhead{($10^{-4}$ cm$^{-3}$)} &
  \colhead{} 
  }  
\startdata

$< 5.81$       & $1.01 \pm 0.06$ & $-$            & $-$             & $0.45 \pm 0.01$ & $6.70 \pm 0.19$ & $6.25 \pm 0.30$ & 2683 (736) \\
$< 5.42$       & $1.35$ (fixed)  & $4.47$         & $2.12 \pm 0.22$ & $0.49 \pm 0.01$ & $6.67 \pm 0.19$ & $6.20 \pm 0.29$ & 2669 (736) \\
$3.83$ (fixed) & $1.35$ (fixed)  & $-$            & $-$             & $0.50$ (fixed)  & $6.61 \pm 0.18$ & $6.01 \pm 0.27$ & 2716 (739) \\
$3.83$ (fixed) & $1.35$ (fixed)  & $2.60$ (fixed) & $3$ (fixed)     & $0.50$ (fixed)  & $7.17 \pm 0.17$ & $7.48 \pm 0.22$ & 2783 (739)

\enddata
\label{table.mcmc_results_chap_fb}

\tablecomments{Unless noted otherwise, all best-fit values are defined as the most likely parameter values from the MCMC marginalized posterior probability distributions.  The uncertainties encompass the 68\% probability region relative to the maximum likelihood value for each parameter.  }

\tablenotetext{a}{The Local Bubble (LB) produces minimal \oviii emission in our model.  With this in mind, we report either the fixed value for $n_{LB}$ or the 2$\sigma$ upper limit from our MCMC analysis.}
\tablenotetext{b}{These best-fit densities assume that each component has a temperature of log($T_{FB,\ shell}$) = 6.50 when calculating line intensities.  See Table~\ref{table.inferred_prop_chap_fb} for densities with higher assumed temperatures.  }

\end{deluxetable*}

\begin{figure}
\begin{center}
\includegraphics[width =.5\textwidth, keepaspectratio=true]{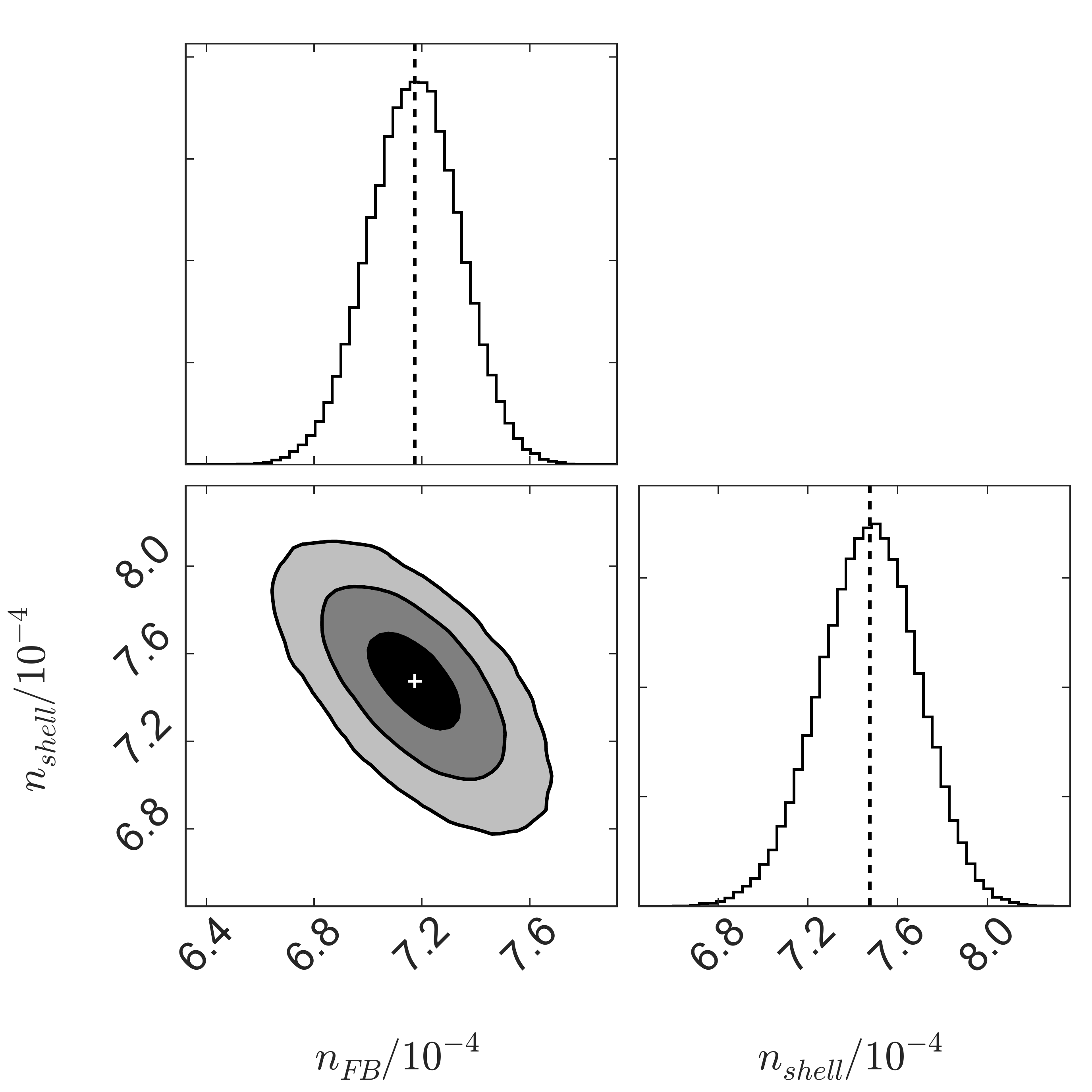}
\caption[Probability distribution functions for the Fermi bubble and shell parameters]{Our model fitting results for the Fermi bubbles' volume-filled and shell components represented as marginalized posterior probability distributions and a two-dimensional contour plot.  The dashed lines and white cross represent the best-fit model values and the contour ranges include 1$\sigma$, 2$\sigma$, and 3$\sigma$.  This model assumes that the bubble and shell components have a temperature of log($T$) = 6.50.   
}
\label{figure.pdf_chap_fb}
\end{center}
\end{figure}

Table~\ref{table.mcmc_results_chap_fb} summarizes our best-fit model parameters, including 1$\sigma$ uncertainties encompassing the 68\% probability ranges from the posterior probability distributions.  There are several trends to note from these results.  The LB density parameter is consistent with zero, validating the assumption that the LB contributes little emission to the \oviii data.  The halo density profile results are consistent with those reported in \citetalias{miller_bregman15} when considering the same power-law density parametrization ($n_or_c^{3\beta} = 1.35 \times 10^{-2}$ cm$^{-3}$ kpc$^{3\beta}$, $\beta$ = 0.5).  This implies that the hot gas density profile constraints are not biased due to observations near the Fermi bubbles.  We also find characteristic best-fit core radii of 2--3 kpc, which is expected.  The parameters of interest, $n_{FB}$ and $n_{shell}$, have characteristic best-fit densities of (5--8)$\times 10^{-4}$ cm$^{-3}$ assuming a temperature of log($T$) = 6.50.  The inferred densities are lower if we assume a power-law model for the halo gas density than if we assume a $\beta$-model.  We expect to see this trend since the power-law model produces more halo emission near the Galactic center than a $\beta$-model with a core radius/density, thus resulting in less Fermi bubble/shell emission being required to produce the observed emission.  After weighing these effects, we define our fiducial model to be one with a $\beta$-model and $r_c$ fixed to 3 kpc.  This results in best-fit parameters of $n_{FB} = 7.2 \pm 0.2 \times 10^{-4}$ cm$^{-3}$, and $n_{shell} = 7.7 \pm 0.2 \times 10^{-4}$ cm$^{-3}$.  Figure~\ref{figure.pdf_chap_fb} shows the marginalized posterior probability distributions and contour plots from our MCMC analysis assuming this parametric model \citep[generated using the Python code \texttt{corner.py}; ][]{foreman_mackey_corner16}.


We also explore models with either the bubble or shell distributions to determine each component's significance in our model fitting procedure.  A bubble-only model is equivalent to setting the shell thickness to 0 kpc and not including $n_{shell}$ as a model parameter parameter.  A shell-only model is the same as our original emission model ($r_c$ fixed to 3 kpc) but with $n_{FB}$ fixed to zero.  When we refit the data with these models, we find that each density component increases to compensate for the lack of emission from the other component.  For example, $n_{FB}$ increased to $7.7\times 10^{-4}$ cm$^{-3}$ in the bubble-only model, and $n_{shell}$ increased to $10.0\times 10^{-4}$ cm$^{-3}$ in the shell-only model.  These best-fit models lead to changes in the overall fit quality, where our initial best-fit $\chi^2_{r}$ (dof) = 2783 (739).  The bubble-only model leads to a marginal improvement in the overall fit quality ($\chi^2_{r,\ bubble-only}$ (dof) = 2741 (740)), while the shell-only model leads to a worse quality of fit ($\chi^2_{r,\ shell-only}$ (dof) = 3241 (740)).  This implies that the volume-filled structure is more important than the shell structure, although we point out that this exercise fixes every other component of the emission model (halo emission, bubble/shell geometry, etc.).  Thus, we still assume that the bubble and shell structures are each present in our discussion and temperature analysis.

In order to constrain the bubble and shell temperatures, we compare best-fit model line ratios for different temperature distributions with the observed \oviii/\ovii line ratios near the Fermi bubbles.  To do this, we change the bubble and shell temperatures while keeping the product $n^2 \epsilon_{OVIII}(T)$ fixed from the best-fit model results.  This fixes the \oviii emission coming from the bubble and shell, but changes the model \ovii emission because $\epsilon_{OVII}(T)$ decreases faster than $\epsilon_{OVIII}(T)$ with increasing temperature.  Thus, increasing the assumed temperature leads to an increase in the model line ratios, an increase in the inferred best-fit densities ($\epsilon_{OVIII}$ decreases for $T > 3 \times 10^6$ K), and a constant contribution to the \oviii emission.  

\begin{figure}
\begin{center}
\includegraphics[width = .5\textwidth, keepaspectratio=true]{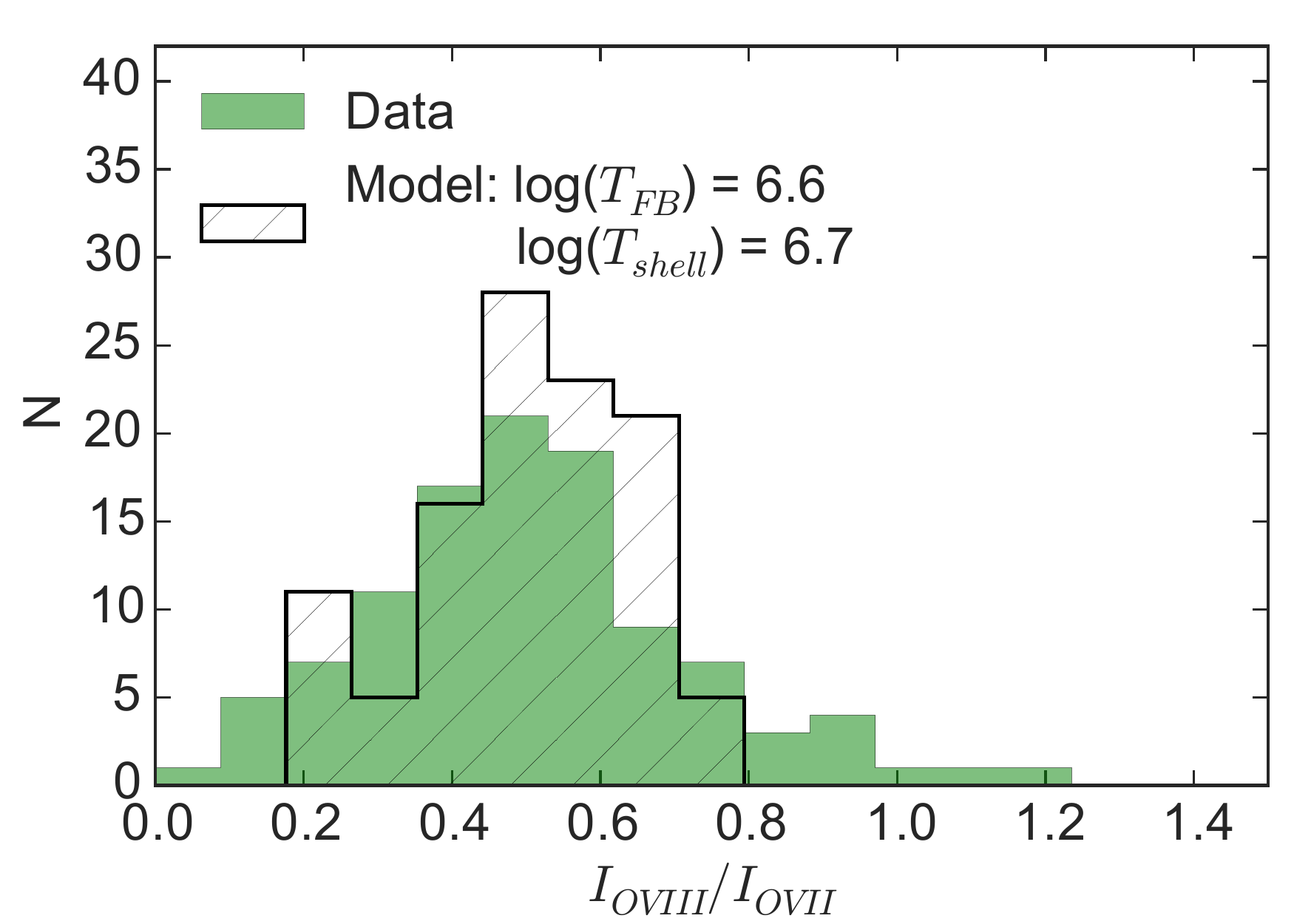}
\caption[Observed and model line ratio histograms near the Fermi bubbles]{Histograms of observed (shaded green areas) and model (black hatched areas) \oviii/\ovii line ratios for sight lines passing through the Fermi bubbles.  The model assumes log($T_{FB}$) = 6.60 and log($T_{shell}$) = 6.70, which produces line ratios that are most consistent with the observations.
}
\label{figure.ratio_hist_chap_fb}
\end{center}
\end{figure}

A model temperature distribution with log($T_{FB}$) = 6.60 and log($T_{shell}$) = 6.70 leads to a model line ratio distribution most consistent with the observed line ratios near the Fermi bubbles.  This changes the inferred best-fit densities to $n_{FB} = 8.2\times 10^{-4}$ cm$^{-3}$ and $n_{shell} = 1.0\times 10^{-3}$ cm$^{-3}$ in order to keep the product $n^2 \epsilon_{OVIII}(T)$ fixed for each component.  Figure~\ref{figure.ratio_hist_chap_fb} shows histograms of the observed and new best-fit model line ratios for sight lines that pass within $\approx 5 \arcdeg$ of the projected bubble edge ($\sim$100 sight lines).  These densities and temperatures produce an \oviii/\ovii ratio median and interquartile range of 0.52 (0.41--0.60), consistent with the observed median and interquartile range of 0.49 (0.38--0.62).  We treat these densities and temperatures as the characteristic physical properties for the bubble and shell components in our subsequent analysis.  


\section{Discussion}
\label{section.discussion_chap_fb}

In this section, we discuss how our constraints fit in with our current picture for the Fermi bubbles and Milky Way.  This includes an overview of the constrained thermal gas structure, and how this compares to the surrounding hot medium.  We extend these constraints to infer the bubbles' characteristic shock strength, current expansion velocity, energy input rate, age, and likely formation scenario.  We also discuss how our constraints compare with previous Fermi bubble analyses.

Table~\ref{table.inferred_prop_chap_fb} summarizes our most important inferred quantities discussed below for the best-fit densities and temperatures discussed above.  The density uncertainties follow directly from our MCMC results summarized in Table~\ref{table.mcmc_results_chap_fb}.  We use less strict criteria for the temperature uncertainties since we did not directly fit the \oviii/\ovii line ratios.  The temperature limits represent where the differences between the observed and model line ratio medians are less than 1$\sigma$ of the corresponding uncertainties in the sample median.  Uncertainties on all subsequent calculated quantities (masses, expansion rates, ages, etc.) use the density and temperature uncertainties listed in Table~\ref{table.inferred_prop_chap_fb}.

\subsection{Inferred Bubble Structure}
\label{subsection.structure_chap_fb}

\begin{figure*}
\begin{center}
\includegraphics[width = 1.\textwidth, keepaspectratio=true]{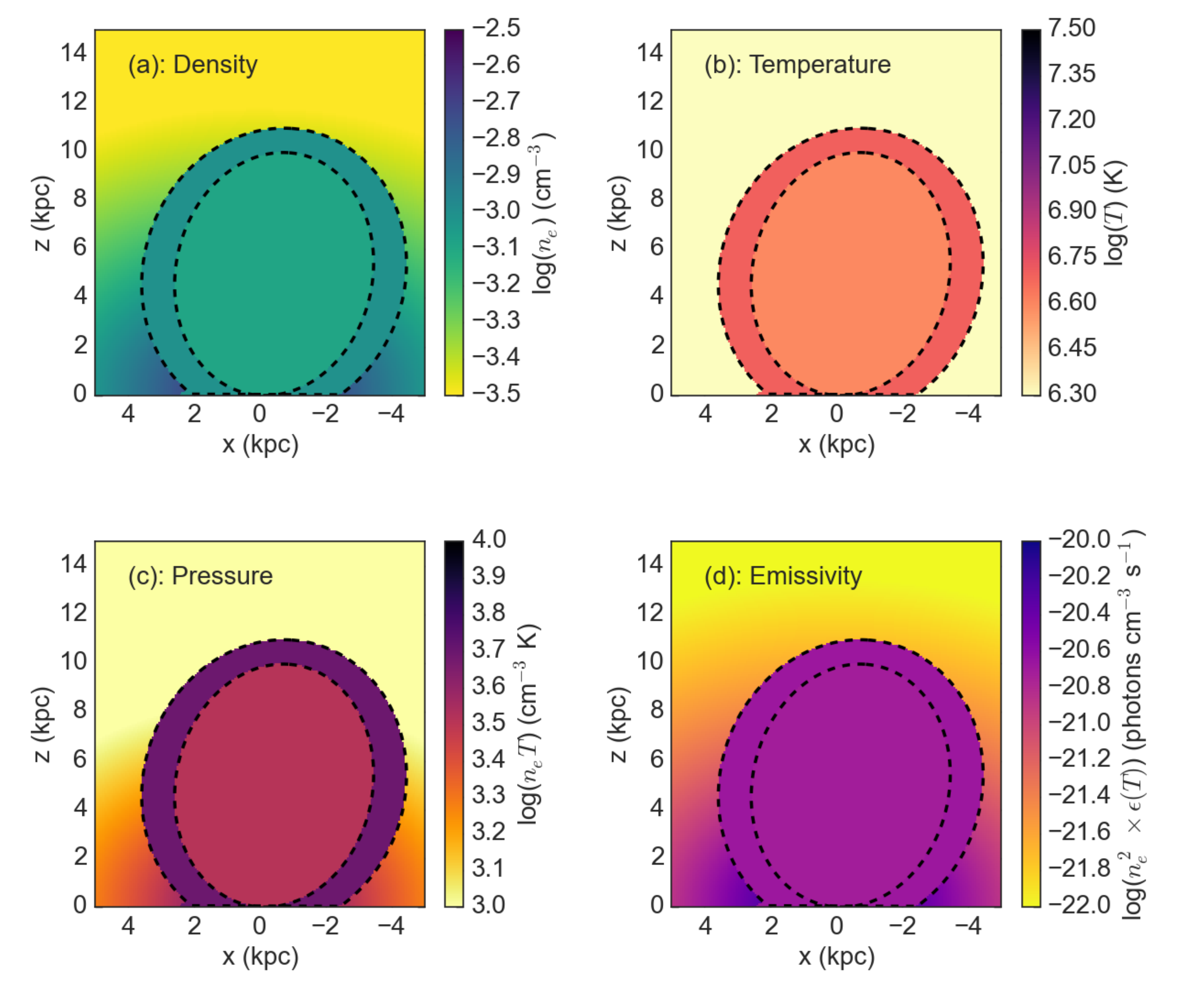}
\caption[Best-fit model profiles as two-dimensional slices]{Our best-fit Fermi bubble model compared to the surrounding hot gas halo profile as two-dimensional slices at the Galactic center.  The dashed lines represent the boundaries between the bubble and shell surfaces.  The panels represent: density ({a}), temperature ({b}), pressure ({c}), and \oviii emissivity defined as $n_e^2 \times \epsilon(T)$ ({d}).  There is variation with height away from the Galactic plane, but the bubbles are hotter and overpressurized compared to the surrounding halo medium.  
}
\label{figure.fit_slice_chap_fb}
\end{center}
\end{figure*}

We discuss our inferred bubble densities and temperatures in this section, and compare them to the assumed ambient structure.  Overall, our constraints indicate that the bubbles are hotter and overpressurized compared to the surrounding medium, consistent with previous observations of the bubble \citep{su_etal10, kataoka_etal15}.  Figure~\ref{figure.fit_slice_chap_fb} shows our best-fit model as two-dimensional maps of density, temperature, and pressure projected at the Galactic center.  This visualizes the comparison with the surrounding medium that we discuss in the rest of the section.

The derived structure indicates that the best-fit densities and temperatures for the bubble and shell components are nearly identical to each other.  This suggests that there may not be two distinct outflow regions, which contradicts the global morphology predicted from Galactic outflow simulations \citep[e.g., ][]{yang_etal12,sarkar_etal15,sofue_etal16}.  This inference could be due to our assumed geometry, in terms of both the volume-filled shape and the shell thickness.  Our emission model effectively constrains each component's emission measure, so changes in the bubble/shell path lengths along different sight lines could change the inferred densities.  Additional substructure could also be present inside the volume-filled component, which might explain why we infer a relatively high density permeating the entire bubble volume.  Modeling this substructure is beyond the scope of this work, but our modeling results still provide valuable constraints on the average bubble and shell properties.


\begin{deluxetable}{l c c c}[t]
\tablewidth{0pt}
\tablecaption{Bubble Properties}
\tablehead{
  \colhead{Quantity} &
  \colhead{Value} &
  \colhead{Uncertainty/Range} &
  \colhead{Unit}
  }  
\startdata

$n_{FB}$      & 8.2  & $\pm 0.2$  & $10^{-4}$ cm$^{-3}$ \\
log($T_{FB}$) & 6.60 & 6.60--6.65 & ($T_{FB}$ in K) \\
$P_{FB}$      & 4.5  & 4.5--5.5   & $10^{-13}$ dyn cm$^{-2}$ \\
$M_{FB}$      & 4.6  & 4.6--5.0   & $10^6$ $M_{\odot}$ \\
\hline
$n_{shell}$      & 10.0 & $\pm 0.3$  & $10^{-4}$ cm$^{-3}$ \\
log($T_{shell}$) & 6.70 & 6.60--6.95 & ($T_{shell}$ in K) \\
$P_{shell}$      & 6.9  & 4.7--19.7  & $10^{-13}$ dyn cm$^{-2}$ \\
$M_{shell}$      & 6.1  & 5.2--9.8   & $10^6$ $M_{\odot}$ \\
\hline
$\mathcal{M}$ & 2.3 & 1.9--3.4 & ... \\
$v_{exp}$     & 490 & 413--720 & km s$^{-1}$ \\
$t_{dyn,\ h}$\tablenotemark{a} & 20.0 & 13.6--23.7 & Myr \\
$t_{dyn,\ w}$\tablenotemark{a} & 6.0  & 4.1--7.11  & Myr \\
$t_{age}$\tablenotemark{b} & 4.3  & 2.9--5.1   & Myr \\
$2 \times \xi \times \dot{E}$\tablenotemark{c} & 2.3 & 1.4--7.4 & $10^{42}$ erg s$^{-1}$

\enddata
\label{table.inferred_prop_chap_fb}
\tablecomments{Summary of our inferred bubble properties discussed in Section~\ref{section.discussion_chap_fb}.  The densities have uncertainties from the MCMC analysis, while the temperatures (and all other derived quantities) have 1$\sigma$ uncertainties based on the difference between the observed and model median line ratio.}

\tablenotetext{a}{$t_{dyn} = d/v_{exp}$, where $t_{dyn,\ h}$ is for the full bubble height and $t_{dyn,\ w}$ is for half the bubble width.}
\tablenotetext{b}{$t_{age}$ is the bubble age defined in Equation~\ref{eq.age_chap_fb}.}
\tablenotetext{c}{$\xi \dot{E}$ is the energy injection rate defined in Equation~\ref{eq.edot_chap_fb}.}

\end{deluxetable}


The bubble and shell densities have characteristic values of $\sim 10^{-3}$ cm$^{-3}$, which are comparable to the surrounding medium at low $z$.  Including a core radius for the hot gas halo of 3 kpc in our fiducial model implies a core density of $2.6 \times 10^{-3}$ cm$^{-3}$, assuming a fixed power-law normalization of $n_o r_c^{3\beta} = 1.35 \times 10^{-2}$ cm$^{-3}$ kpc$^{3\beta}$.  This suggests $n_{shell} \sim n_{halo}$ within $|z| \lesssim $5 kpc.  The hot gas halo density decreases by about a factor of 6 between $r = $1 and 10 kpc, meaning that our bubble and shell densities are larger than the surrounding CGM density farther away from the Galactic plane.  We also note that $n_{FB} \approx n_{shell}$ from our model fitting results, making it difficult to distinguish between volume-filling emission and limb-brightened emission.  This might be due to our choice to parameterize the structures with constant densities and temperatures, but our constraints still probe the average densities associated with the bubbles.  

Our inferred bubble and shell temperatures of log($T_{FB,\ shell}$) = 6.60--6.70 are hotter than the surrounding medium (log($T_{halo}$) = 6.30).  This is broadly consistent with the bubbles injecting enough energy to shock-heat the surrounding medium, although simulations predict a wide range of shock strengths and bubble temperatures as high as $\sim 10^8$ K \citep[e.g., ][]{guo_mathews12a, yang_etal12}.  It is possible that the bubbles contain gas at this high temperature, but plasma at this temperature would not produce observable \oviii emission.  Our modeling results indicate that on average the bubble and shell are hotter than the surrounding medium, but still at low enough temperatures to produce observable signatures in the data.  

Combining the density and temperature constraints indicates that the bubbles' are overpressurized compared to the surrounding medium.  Our best-fit bubble and shell parameters indicate thermal gas pressures of $P_{FB} = 4.5 \times 10^{-13}$ dyn cm$^{-2}$, $P_{shell} = 6.9 \times 10^{-13}$ dyn cm$^{-2}$, or $P/k \approx $3000--5000 cm$^{-3}$ K.  The surrounding thermal gas pressure varies with $r$ and $z$ due to the decreasing density profile, with characteristic values of $\approx$5000 cm$^{-3}$ K near the Galactic center and $\approx$1000 cm$^{-3}$ K at $r$ = 10 kpc.  In this picture, the bubbles are in approximate pressure equilibrium at lower $z$, but become overpressurized with increasing height away from the Galactic plane.  These estimates are also a lower limit to how overpressurized the bubbles actually are, because they do not account for non-thermal or magnetic pressure contributions.  Nevertheless, these constraints indicate that the bubbles are generally overpressurized, and thus expanding into the surrounding medium.  

We use these quantities to infer a characteristic shock strength and instantaneous expansion velocity associated with the bubbles.  The classic treatment of shocks propagating through the ISM yields specific pre- and post-shock jump conditions for the gas density, temperature, and pressure given a shock expansion velocity (e.g., Shull \& Draine \citeyear{shull_draine87}, pp. 283--319).  The Fermi bubbles' expansion is more complex than this traditional treatment since they do not appear to be spherical, and they are presumably expanding into a medium with varying density.  For example, the ratio between $n_{shell}$ (treated as post-shocked material) and the ambient halo gas density along the shell edge (treated as pre-shock material) ranges between $\approx$0.5 near the Galactic center and $\approx$3 at the maximum bubble height.  On the other hand, our choice to parameterize the hot gas halo and shell with constant temperatures allows us to use the temperature ratio as a shock strength diagnostic similar to \citetalias{kataoka_etal15}.  Assuming a monotonic gas ($\gamma$=5/3), an ambient gas sound speed of $c_s$= 212 km s$^{-1}$, $T_{shell} = 5 \times 10^6$ K, and $T_{halo} = 2 \times 10^6$ K, our temperature ratio implies a fiducial Mach number and corresponding expansion velocity of $\mathcal{M} = 2.3$ and $v_{exp}$ = 490 km s$^{-1}$.  The uncertainty in $T_{shell}$ expands these constraints to $\mathcal{M} =$1.9--3.4 and $v_{exp}$ = 413--720 km s$^{-1}$.  These shock parameters are broadly consistent with the range of density and pressure ratios we estimate, indicating that these constraints probe the bubbles' current expansion rate into the surrounding medium. 


\subsection{Bubble Energetics and Origin Scenarios}
\label{subsection.origins_chap_fb}

\subsubsection{The Bubbles as a Confined Galactic Wind}
\label{subsubsection.wind_chap_fb}

We treat the bubbles in the framework of a continuous galactic outflow/superbubble with self-similar Sedov-Taylor solutions \citep[e.g., ][]{castor_etal75, weaver_etal77, maclow_mccray88, veilleux_etal05}.  The outflow morphology consists of five zones (from closer to farther from the outflow origin): the energy injection zone, a free-flowing outflow, shocked wind material, a shell of shocked ISM/CGM material, and the ambient ISM/CGM.  Our model constraints probe the last three zones since we do not model observations in the inner $\approx$1 kpc from the Galactic center.  The Sedov-Taylor solutions for this type of outflow relate the ambient density, bubble age, bubble size, expansion velocity, and average energy injection relate to each other.  Assuming that the outflow is still in the energy-conserving phase (cooling time greater than the bubble age), the relations between these quantities are as follows: 

\begin{equation}
\label{eq.age_chap_fb}
t_{age}
=\ 11.8\ Myr\ 
\left(\frac{r}{10\ \text{kpc}}\right)
\left(\frac{v}{500\ \text{km s}^{-1}}\right)^{-1}, 
\end{equation}

\begin{equation}
\label{eq.edot_chap_fb}
\begin{aligned}
&\xi\dot{E}
=\ 3.7 \times 10^{42}\ \text{erg s}^{-1}\ 
\\
&\times \left(\frac{n_o}{10^{-3}\ \text{cm}^{-3}}\right)
\left(\frac{r}{10\ \text{kpc}}\right)^2
\left(\frac{v_{exp}}{500\ \text{km s}^{-1}}\right)^3, 
\end{aligned}
\end{equation}

\noindent where $n_o$ is the ambient density, $r$ is the bubble radius, $v_{exp}$ is the expansion velocity, $t_{age}$ is the bubble age, $\dot{E}$ is the energy injection rate, and $\xi$ is the thermalization efficiency of the mechanical energy.  This thermalization efficiency is believed to vary with environment, but is estimated to be $\gtrsim$10\% in average galaxies with a typical assumed value of 0.3 \citep[e.g., ][]{larson74, wada_norman01, melioli_degouvia04}.

Our modeling results constrain three of the five variables in these equations.  As discussed above, a hot gas halo model with $r_c$ = 3 kpc and fixed power-law normalization of $1.35 \times 10^{-2}$ cm$^{-3}$ kpc$^{3\beta}$ results in an ambient core density of $n_o = 2.6 \times 10^{-3}$ cm$^{-3}$.  The constraints on the bubble and shell temperatures suggest an outflow velocity of 490 km s$^{-1}$.  The bubble size is not trivial to estimate in this framework, where the outflow is typically treated as a spherical shell with radius $r$ centered on the injection source.  The bubbles' shape is more complex than this since two lobes exist on each side of the Galactic plane.  We use a characteristic bubble size defined as the geometric mean of the three ellipsoidal axes, resulting in $r$ = 3.6 kpc.

Given these constrained values, we estimate the bubbles' age and average mechanical energy injection rate.  The bubbles' dynamical timescale, $t_{dyn} = d/v_{exp}$, is a crude age estimate that does not incorporate the bubble environment or energy source.  For $v_{exp}$ = 490 km s$^{-1}$, the dynamical timescale for the full bubble height is $t_{dyn,\ h}$ = 10 kpc / 490 km s$^{-1}$ = 20.0 Myr, and the dynamical timescale for half the bubble width is $t_{dyn,\ w}$ = 3 kpc / 490 km s$^{-1}$ = 6.0 Myr.  The superbubble model calculation (Equation~\ref{eq.age_chap_fb}) is a refined age estimate, where we find $t_{age}$ = 4.3 Myr for $r$ defined as the geometric mean above and $v_{exp}$ = 490 km s$^{-1}$.  We also infer a combined energy injection for both bubbles ($2 \times \xi \times \dot{E}$) of $2.3 \times 10^{42}$ erg s$^{-1}$ using Equation~\ref{eq.edot_chap_fb}.  Accounting for the uncertainty in $v_{exp}$ leads to a characteristic age range of $\approx$3--5 Myr and energy injection rate of $\approx$1--7$ \times 10^{42}$ erg s$^{-1}$.  We compare this characteristic age and energy injection rate with possible bubble formation mechanisms.  

\subsubsection{Origin from Sgr A* Accretion}
\label{subsubsection.sgra_chap_fb}

One suggested bubble formation mechanism has been a past accretion event onto Sgr A*, resulting in an AGN episode in the Milky Way.  Sgr A* has an estimated mass of $4 \times 10^6\ M_{\odot}$ \citep{schodel_etal02,ghez_etal03,ghez_etal08,gillessen_etal09b,gillessen_etal09a,meyer_etal12}, which is capable of producing significant amounts of energy during an accretion episode.  We also know that accretion onto supermassive black holes can produce galactic outflows with significant energy injection rates and morphologies similar to the observed Fermi bubbles \citep[e.g., ][]{mcnamara_nulsen07,yuan_narayan14}.  Here, we consider observations of Sgr A* and its possible accretion history, and compare the expected energetics with our modeling constraints.

Sgr A* is currently in a quiescent state with a bolometric luminosity of $L_{bol} \sim 10^{36}$ erg s$^{-1}$ $\sim 2 \times 10^{-9}\ L_{Edd}$ \citep[e.g., ][]{yuan_narayan14}.  Our proximity to Sgr A* allows for a combination of techniques to estimate current mass accretion rates.  $Chandra's$ resolution is comparable to the Sgr A* Bondi radius, and has constrained the Bondi accretion rate to be $\sim 10^{-5}$ $M_{\odot}$ yr$^{-1}$ \citep{baganoff_etal03}.  Polarized radio emission constrains the accretion rate near the event horizon, with limits being between $> 2 \times 10^{-9}$ $M_{\odot}$ yr$^{-1}$ and $< 2 \times 10^{-7}$ $M_{\odot}$ yr$^{-1}$ depending on the magnetic field orientation \citep[e.g., ][]{marrone_etal07}.  While this is a significant uncertainty in the current mass accretion rate, the consensus is that Sgr A* is accreting well below its Eddington rate, and has been a well-modeled source for radiatively inefficient accretion flows (RIAFs).  

There are a number of observational indications that Sgr A* has been more active in the past \citep{totani06}.  \citet{mou_etal14} summarizes these lines of evidence, which include: a higher Sgr A* luminosity is required to produce fluorescent iron emission and reflection nebulae seen in several nearby molecular clouds \citep{koyama_etal96,murakami_etal00,murakami_etal01b,murakami_etal01a}, there exists an ionized halo of material surrounding Sgr A* \citep{maeda_etal02}, there are dynamic features indicating an outflow near the Galactic center in the form of the Galactic Center Lobe \citep{bh_cohen03} and the Expanding Molecular Ring \citep{kaifu_etal72, scoville72}, excess H$\alpha$ emission seen in the Magellanic Stream \citep{bh_etal13}, and possibly the Fermi bubbles themselves.  The RIAF modeling from \citet{totani06} argues that Sgr A* should have had an accretion rate $\sim 10^3$--$10^4$ times larger than its current accretion rate over the past $\sim$10 Myr to reproduce these observations.  This introduces additional scatter in the inferred past Sgr A* accretion rate, but motivates the assumption that Sgr A* has injected energy into the surrounding medium through an accretion event.  

We estimate an energy injection rate due to past Sgr A* accretion and compare with our constrained energy input rate.  The mechanical energy injection rate from black hole accretion is tied to the accretion power by the following relation:

\begin{equation}
\label{eq.bh_chap_fb}
\begin{aligned}
\dot{E}_{BH} &= \epsilon \dot{M}_{acc}c^2
\\
&= 5.7 \times 10^{45}\ \text{erg s}^{-1}\ 
\left( \frac{\epsilon}{0.1} \right) \left(\frac{\dot{M}_{acc}}{M_{\odot}\ \text{yr}^{-1}}\right), 
\end{aligned}
\end{equation}

\noindent where $\dot{E}_{BH}$ is the mechanical energy injection rate, $\dot{M}_{acc}$ is the accretion rate near the event horizon, and $\epsilon$ is the mechanical energy injection rate efficiency.  If we assume a past accretion rate of $10^{-3}$ $M_{\odot}$ yr$^{-1}$ (near the high end of the values discussed above), we find that $\dot{E}_{BH}$ equals our inferred vale of $2.3 \times 10^{42}$ erg s$^{-1}$ for $\epsilon \approx 0.05$.  This efficiency is larger than the typical values inferred from simulations \citep[$10^{-4}$--$10^{-3}$; ][]{yuan_etal15}, but this mechanical efficiency is often treated as a free parameter in simulations.  We also point out that the required efficiency is less than one, indicating that this analysis does not violate energy conservation constraints.  \textit{Thus, it is plausible that a past accretion episode onto Sgr A* could have produced enough energy to match our energy injection rate constraints.}

The bubble age indicates that this Sgr A* accretion episode had a shorter active period than the typical AGN duty cycle.  Studies constrain the AGN duty cycle by either comparing black hole mass functions (inferred from the $M_{BH}$--$\sigma$ relation) to AGN luminosity functions at different redshifts \citep[e.g., ][]{yu_tremaine02,shankar_etal04,hao_etal05,schawinski_etal10} or through analytic models of black hole accretion \citep[e.g., ][]{hopkins_hernquist06,shankar_etal09}.  These techniques suggest that black holes with masses of $\sim 10^6$ \msun should have active periods of $\sim 10^8$ yr at $z$ = 0, or $\sim$1\% of a Hubble time.  Our Fermi bubble age estimate is an upper limit to the active Sgr A* accretion time, and our constraint of 4.3 Myr is much smaller than the inferred duty cycle from AGN populations.  One possible explanation for this discrepancy is that the Fermi bubble outburst was one of several past accretion events in the Galactic center.  Our constraints imply that the Fermi bubbles are due to a relatively weak AGN event, and it is possible that multiple Sgr A* accretion events of comparable or lower energy have occurred over the past $\sim 10^8$ yr.  Our results are also consistent with the overall decrease in AGN activity since $z \sim 2$ \citep[e.g., ][]{hopkins_etal07}, as opposed to a prolonged Sgr A* accretion/growth phase.

\subsubsection{Origin from Nuclear Star Formation}
\label{subsubsection.nsf_chap_fb}

Numerous studies also suggest that the Fermi bubbles formed from a period of enhanced star formation activity near the Galactic center.  The Galactic center hosts several young stellar clusters with ages ranging between 5 and 20 Myr and accounting for $\sim 5 \times 10^5$ $M_{\odot}$ of material.  The massive stars in these clusters could have generated a galactic-scale outflow due to stellar winds and type-II supernova explosions \citep[e.g., ][]{leitherer_etal99}.  Here, we compare the expected energy output from past star formation near the Galactic center to our energy injection rate constraints.  

The Galactic center star formation history is complex and difficult to measure, but several studies argue for an average star formation rate (SFR) of $\approx 0.05$ $M_{\odot}$ yr$^{-1}$ over the past $\sim$10 Myr.  \citet{crocker12} reviews these studies, most of which utilize \textit{Spitzer} observations of young stellar objects within the inner $\sim$500 pc from the Galactic center.  For example, \citet{yusefzadeh_etal09} conducted a census of these objects using the Infrared Array Camera and Multiband Imaging Photometer on board \textit{Spitzer}, and concluded the average SFR has been between 0.04 and 0.08 $M_{\odot}$ yr$^{-1}$ over longer timescales ($\sim$10 Gyr).  \citet{immer_etal12} performed a similar analysis using data from the \textit{Spitzer} Infrared Spectrograph, and argue for an average SFR of $\approx 0.08$ $M_{\odot}$ yr$^{-1}$ over the past $\sim$Myr.  Others estimate the SFR to be $\approx$0.01--0.02 $M_{\odot}$ yr$^{-1}$ by counting the mass in young star clusters and dividing that by estimates for the period of star formation \citep{figer_etal04, mauerhan_etal10}.  Thus, it appears that a characteristic SFR of $\approx 0.05$ $M_{\odot}$ yr$^{-1}$ over the past $\sim$10 Myr is a reasonable assumption.  

Similar to our argument concerning the black hole accretion energy above, we estimate the energy injection rate due to star formation in the Galactic center to compare with our constrained energy input rate.  Assuming a Kroupa initial mass function \citep{kroupa01}, and $10^{51}$ erg of mechanical energy input from a type-II supernova, the mechanical energy from type-II supernovae is related to the SFR as

\begin{equation}
\label{eq.nsf_chap_fb}
\dot{E}_{nsf} = 1.1\ \times 10^{40}\ \text{erg s}^{-1}\ \left( \frac{\epsilon}{0.3} \right) \left(\frac{SFR}{0.1\ M_{\odot}\ \text{yr}^{-1}}\right), 
\end{equation}

\noindent where $\dot{E}_{nsf}$ is the mechanical energy input rate due to nuclear star formation and $\epsilon$ is an efficiency factor typically assumed to be $\approx$0.3 (see Crocker et al. \citeyear{crocker_etal15} or Sarkar et al. \citeyear{sarkar_etal15} for equivalent relations).  This implies that an average SFR of 0.05 $M_{\odot}$ yr$^{-1}$ over the past $\sim$10 Myr produces an energy injection rate of $\approx 6 \times 10^{39}$ erg s$^{-1}$.  This estimate falls $\approx$400 times lower than our estimated energy input rate of $2.3 \times 10^{42}$ erg s$^{-1}$.  It is possible that the SFR has been more variable over the past $\sim$10 Myr, however the upper limits are only $\approx$3 times higher than the average value \citep{yusefzadeh_etal09}.  \textit{Thus, star formation in the Galactic center does not produce enough energy to inflate the bubbles based on our energy injection rate constraints.}

\subsection{Thermal Gas Masses}
\label{subsection.mass_chap_fb}

We use our density constraints to estimate the thermal gas mass within the bubble and shell structures.  This is a straightforward calculation since we assume that each component has a constant density and fixed volume.  Thus, the mass in each component is defined as $M = \mu m_H \times n \times V$, where $\mu = 0.61$ is the average weight per particle, $m_H$ is the mass of hydrogen, $n$ is the inferred density, and $V$ is the volume.  Our geometric models imply a bubble volume of $V_{FB} = 2 \times 4/3 \times \pi \times 5 \times 3^2 = 377$ kpc$^3$ (the factor of two is for two ellipsoidal bubbles), and a combined shell volume of $V_{shell} = 411$ kpc$^3$.  The densities in Table~\ref{table.inferred_prop_chap_fb} imply thermal gas masses of $M_{FB} = 4.6 \times 10^6$ $M_{\odot}$ and $M_{shell} = 6.1 \times 10^6$ $M_{\odot}$ for the bubble and shell, with a characteristic range between 5 and 10$\times 10^6$ $M_{\odot}$ given the density uncertainties.  These masses represent material that has been shock-heated by the bubbles or injected into the bubbles by the energy source.

We first explore whether the bubble and shell plasmas are predominantly shocked/mixed hot halo material by comparing the masses derived above to the inferred hot gas halo mass that would exist within the bubble and shell volumes.  If the density inside the bubble+shell volume was defined by our hot gas halo density model with  $r_c$ = 3 kpc (instead of the Fermi bubble/shell densities), the halo mass in the volume would be $1.11 \times 10^7$ $M_{\odot}$.  The calculations above indicate that the combined thermal gas mass within the bubble+shell volumes is $M_{FB}+M_{shell} = 1.07 \times 10^7$ $M_{\odot}$.  The consistency between these values suggests that most of the thermal gas associated with the bubbles is shock-heated ambient material, as opposed to material injected by the energy source.  


\begin{figure}[t]
\begin{center}
\includegraphics[width = .5\textwidth, keepaspectratio=true]{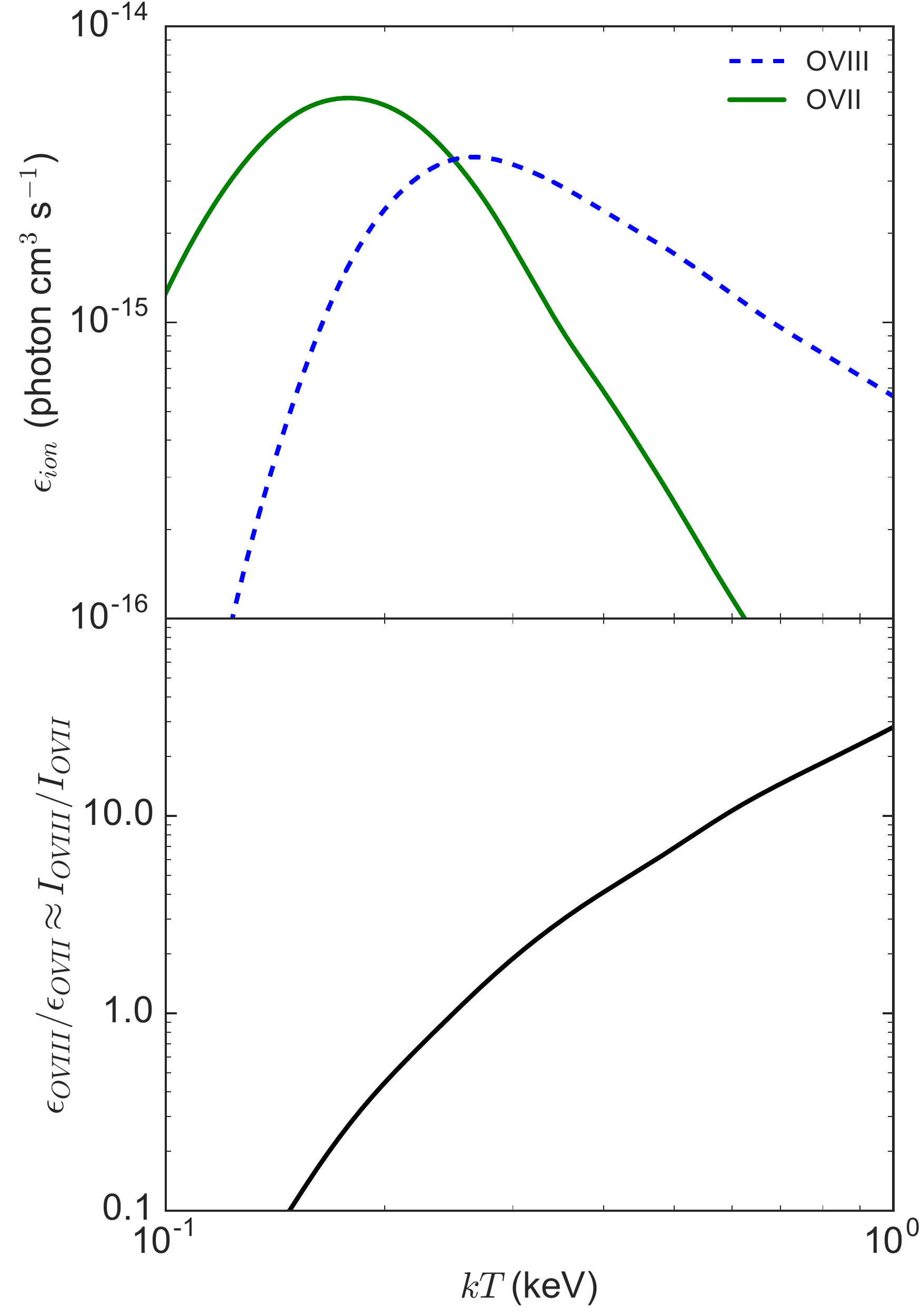}
\caption[Oxygen line emissivities as a function of plasma temperature]{Oxygen line emissivities as a function of plasma temperature.  The {top} panel shows the \ovii (solid green) and \oviii (dashed blue) line emissivities for different APEC plasma temperatures \citep{foster_etal12}.  The {bottom} panel shows the emissivity ratio, which equals the total observed line ratio if the emission consists of a single plasma.
}
\label{figure.epsilon_ratio}
\end{center}
\end{figure}


As a consistency check, we estimate the amount of material injected by the energy source, either AGN or star formation, to compare with the above masses.  The mass-loss rate due to nuclear star formation activity (or mass injection rate) is believed to be $\dot{M}_{inj,\ nsf} \approx 0.3 (SFR / M_{\odot}yr^{-1})$ \citep{leitherer_etal99}.  Mass-loss rates from black hole accretion events (from either jets or winds) are more uncertain, but simulations of RIAF accretion winds suggest values ranging between 2\% and 20\% of $\dot{M}_{Edd} = 10L_{Edd}/c^2$ \citep{yuan_etal12, yuan_etal15}.  Assuming a nuclear SFR of 0.05 $M_{\odot}$ yr$^{-1}$, $\dot{M}_{Edd} \sim 10^{-1}$ $M_{\odot}$ yr$^{-1}$ for Sgr A*, and an active period of for the bubble of 4.3 Myr, we estimate the injected mass is $M_{inj} \lesssim 10^{5}$ $M_{\odot}$ for both origin scenarios.  This is significantly less than our constrained mass estimate of $\sim 10^7$ $M_{\odot}$, thus validating our claim that the bubbles contain predominantly shocked halo gas.

\subsection{Comparing with Previous Work}
\label{subsection.compare_chap_fb}

\subsubsection{Analyses at Soft X-Ray Energies}
\label{subsubsection.softxray_chap_fb}

The most direct comparisons to our analysis are the previous soft X-ray spectral analyses \citep{kataoka_etal13,kataoka_etal15,tahara_etal15}.  These studies follow a similar methodology and find similar results, with \citetalias{kataoka_etal15} being the most current and comprehensive work of the three.  These authors compiled a sample of 29 \textit{Suzaku} observations and 68 \textit{Swift} observations distributed across the Fermi bubbles.  They fit the \textit{Suzaku} XIS data and \textit{Swift} X-ray Telescope spectra with a multi-component thermal plasma model, where one component is typically fixed at $kT = 0.1$ keV to represent emission from the LB and residual SWCX, and the other represents the combined emission from the halo and Fermi bubbles.  They systematically find a hot gas halo/Fermi bubble plasma temperature of $kT = 0.3$ keV, and this is hotter than the characteristic value found for sight lines away from the bubbles \citep[$kT = 0.2$ keV; ][]{hs13}.  From this temperature, they infer a relatively low Mach number of $\mathcal{M}$ = 0.3 keV / 0.2 keV = 1.5 and an expansion velocity of $v_{exp} = 300$ km s$^{-1}$.  They also find emission measures that vary by over an order of magnitude, which they note is due to a combination of emission from the Galactic halo and from the Fermi bubbles.  The authors model the emission measures in the northern Galactic hemisphere with a hot gas halo density model from \citet{miller_bregman13} and a Fermi bubble shell distribution with an inner radius of 3 kpc, an outer radius of 5 kpc, and a density of $3.4 \times 10^{-3}$ cm$^{-3}$.

While our approach is similar to these studies, there are several differences that can explain why we derive a higher bubble temperature and lower bubble density.  The main observational difference is that these studies fit the SXRB spectra for emission measures and temperatures, while we chose to measure oxygen emission line intensities.  Emission measures and temperatures are more useful plasma properties to measure, but fitting a full 0.5--2.0 keV SXRB spectrum with a thermal plasma model requires more counts than fitting only the oxygen emission lines.  This is why our sample is larger and has better sky coverage than the \citetalias{kataoka_etal15} sample.  However, these observables should give consistent results for the inferred bubble temperature and density.  The measured plasma temperature is most sensitive to the \oviii / \ovii ratio for temperatures between $\approx$0.1 and 0.3 keV, because the lines are strong and rapidly changing in strength in this regime (Figure~\ref{figure.epsilon_ratio}).  Thus, any temperature derived from fitting an SXRB spectrum with an APEC model should be consistent with the temperature inferred from fitting the oxygen lines separately \citep{yoshino_etal09}.  The analysis becomes more complicated when there are multiple emission components, and the interpretation of multiple SXRB sources is likely the bigger difference between approaches.

The primary difference between our work and those discussed above is the treatment of combined X-ray emission from the hot gas halo and Fermi bubbles. The observed emission includes contributions from the hot halo and Fermi bubbles.  Thus, the fitted plasma temperature of 0.3 keV is the emission measure-weighted temperature of the hot halo at 0.2 keV and a Fermi bubble plasma that is likely $>$0.3 keV.  The \citetalias{kataoka_etal15} analysis assumes that the Fermi bubbles dominate the observed emission, while our analysis includes the combined emission from the hot halo and Fermi bubbles.  This extension leads to a similar result to these previous works, but also explains why we infer a higher temperature for the Fermi bubble plasma than these studies ($kT_{FB,\ shell} \approx$ 0.4--0.5 keV).

A similar interpretation likely explains why the \citetalias{kataoka_etal15} analysis infers bubble densities 3--4 times higher than our constraints.  Their Fermi bubble geometric distribution includes only a shell component that is 2 kpc thick, while ours includes both a volume-filled component and a shell component.  This implies that our bubble+shell emission model has a longer path length along most sight lines near the bubbles than their model.  The emission measure scales with density and path length as $EM \propto n^2 L$, so a longer inferred path length would lead to a lower inferred density.  We also point out that their hot gas halo density model extends only to $r$ = 20 kpc.  While the hot halo emission is likely dominated by gas within $r \lesssim$25 kpc, failing to account for emission at greater radii can decrease the amount of modeled halo emission.  The combined effect is that the \citetalias{kataoka_etal15} analysis assumes shorter bubble path lengths and less halo emission than our emission model, and this results in a higher inferred bubble density required to match the total observed emission.




\subsubsection{Kinematic Estimates from UV Absorption Lines}
\label{subsubsection.uv_chap_fb}

A different approach to constrain the Fermi bubble kinematics involves analysis of UV absorption lines near background quasars.  \citet{fox_etal15} observed the quasar PDS 456 ($l,b = 10.4\arcdeg, 11.2\arcdeg$) with the Cosmic Origins Spectrograph on board the $Hubble\ Space\ Telescope$.  The spectrum from 1133--1778 \AA\ covers several ionic species indicative of gas with $T \sim 10^4$--$10^5$ K, including \ion{Si}{2}, \ion{Si}{3}, \ion{Si}{4}, \ion{C}{2}, \ion{C}{4}, and \ion{N}{5}.  They detected multiple absorption components for each species, but they argue the nearly symmetric components at $v_{LSR}$ = -235 km s$^{-1}$ and +250 km s$^{-1}$ are unlikely to come from absorbers in the disk or farther in the halo.  If these absorbers represent gas entrained near the bubble edges, their velocities can be used to constrain the bubble kinematics.  Indeed, the authors apply a Galactic wind model from \citet{bordoloi_etal14} to simulate $v_{LSR}$ absorbers and find that an intrinsic outflow velocity of $\geq$900 km s$^{-1}$ is required to reproduce the observed absorption features.  

There is tension between these results and our lower inferred expansion rate of $\approx$500 km s$^{-1}$.  Although we infer a higher expansion rate than \citet{kataoka_etal15}, they discuss this discrepancy as well.  The outflow model used by \citet{fox_etal15} has two important parameters---the outflow velocity and the opening angle.  They assume an opening angle of 110$\arcdeg$ to match the hard X-ray arcs seen by \citet{bh_cohen03}.  However, this geometry produces a significant correction between the intrinsic outflow velocity and $v_{LSR}$ at low latitudes.  Their model implies that most of the bubble velocity at $l,b \approx 10\arcdeg, 10\arcdeg$ is tangential to the line of sight, which may not be the case.  If instead the bubbles have a rounder surface at lower $z$ or a stronger outflow velocity vector away from the Galaxy's polar axis, a lower intrinsic velocity could reproduce the observed absorption.  Thus, the unknown intrinsic bubble geometry plausibly accounts for the different expansion velocities inferred from these two methodologies.

\subsubsection{Comparing with Simulations}
\label{subsubsection.sims_chap_fb}

The Fermi bubbles have motivated numerous simulations of Galactic outflows since their discovery.  Typically, these studies primarily focus on the gamma-ray source, which is tied to the underlying cosmic-ray composition (leptonic or hadronic) and where the cosmic rays are produced (injected from the central source, accelerated in situ, etc.).  All of these simulations predict various distributions for the non-thermal and thermal gas within the bubbles, but information on the latter is often not discussed in detail.  This limits our comparison to characteristic densities, velocities, and energetics, although our results are initial steps toward constraining these properties.  

Simulations are also generally segregated by the assumed energy source, either a black hole accretion event or nuclear star formation.  There is much variation with the assumed outflow parameters, but black hole accretion simulations tend to be more energetic on shorter time scales than star formation simulations.  For example, simulations producing the bubbles with AGN jets have characteristic total energy injection rates and ages of $\gtrsim 10^{44}$ erg s$^{-1}$ and $\approx$1--3 Myr \citep[e.g., ][]{guo_mathews12a,guo_mathews12b,yang_etal12,yang_etal13}, whereas simulations producing the bubbles from weaker AGN winds suggest values of $10^{41}$--$10^{42}$ erg s$^{-1}$ and 5--10 Myr \citep[e.g., ][]{mou_etal14,mou_etal15}.  Alternatively, nuclear star formation simulations can reproduce the bubble morphology with energy injection rates of $\approx$(1--5)$\times 10^{40}$ erg s$^{-1}$ over $\gtrsim$50 Myr timescales \citep[e.g., ][]{crocker_etal14,crocker_etal15,sarkar_etal15}.  The AGNs simulations also tend to predict stronger outflow velocities than the star formation simulations  ($\gtrsim$1000 km s$^{-1}$ compared to $\lesssim$500 km s$^{-1}$).  

Our inferred energy injection rate, bubble age, and expansion velocity are most consistent with the weaker black hole accretion simulations, where the bubbles are inflated by an AGN wind \citep{mou_etal14,mou_etal15}.  Simulations of AGN jets predict higher energy input rates than our results, while star formation simulations are typically weaker and over a much longer timescale than our constraints.  It is difficult to make stronger claims at this point since these simulations are subject to a number of uncertainties.  For example, the energy injection rate required to match the bubble morphology is degenerate with the density of the surrounding medium since it opposes the ram pressure from the galactic wind.  Most simulations assume an ambient density comparable to our core density ($\sim 10^{-3}$ cm$^{-3}$), but this is a well-documented degeneracy in the simulations.  Regardless of these limitations, our constraints should be used to motivate future simulations designed to analyze the bubbles.  

Our thermal pressure constraints also address the bubbles' cosmic-ray composition and whether thermal or non-thermal pressure drives the bubbles' expansion.  Simulations can produce the bubbles' gamma-ray and microwave emission by accelerating either leptonic or hadronic cosmic rays, leading to uncertainties in the inferred non-thermal pressure (cosmic rays and magnetic fields).  For example, the leptonic AGN jet simulations from \citet{yang_etal13} predict a total pressure inside the bubbles of $\sim 10^{-10}$ dyn cm$^{-2}$ and a cosmic ray pressure of $\sim 10^{-12}$ dyn cm$^{-2}$.  This implies either that the bubbles are dominated by a thermal gas pressure much larger than our estimates (magnetic pressure is negligible) or that an additional hadronic cosmic ray source could contribute most of the pressure.  The former scenario is consistent with nuclear star formation simulations that accelerate cosmic-ray leptons \citep{sarkar_etal15} or hadrons \citep{crocker_etal15}.  Alternatively, limits from hard X-ray spectra near the bubbles imply a cosmic ray electron and magnetic pressure of $\approx 2 \times 10^{-12}$ dyn cm$^{-2}$ \citep{kataoka_etal13}, which is approximately equal to the \citepalias{kataoka_etal15} thermal pressure estimate.  Our characteristic thermal gas pressure of (5--20)$\times 10^{-13}$ dyn cm$^{-2}$ should be used in future modeling work to build a more accurate census of the bubbles' energy and pressure budget.




\section{Conclusions}
\label{section.conclusions_chap_fb}

This work is a comprehensive observational analysis of the Fermi bubbles at soft X-ray energies.  The \ovii and \oviii emission line sample includes data from \textit{XMM-Newton} and \textit{Suzaku}, with 741 sight lines in total and $\sim 100$ sight lines projected near the Fermi bubbles.  The new \textit{Suzaku} measurements were processed in a similar way to the \textit{XMM-Newton} measurements, making this the largest emission line sample designed to probe Galactic-scale hot gas distributions.  

We used this sample to model the Fermi bubbles' thermal gas emission, resulting in improved constraints on the bubbles' physical properties and their role in the Milky Way's evolution.  Our modeling procedure is similar to previous studies at soft X-ray energies \citep{kataoka_etal13,kataoka_etal15,tahara_etal15}, although we model the combined emission from the hot halo and Fermi bubbles simultaneously.  This extension confirms the result that the bubbles are hotter than the surrounding medium and expanding supersonically, and leads to a stronger shock than previous works.  Thus, these are improved constraints on the Fermi bubbles' thermal gas distribution given the data currently available.

We summarize our primary conclusions and inferred bubble properties:

\begin{enumerate}

\item{The observed \oviii/\ovii ratios are systematically larger for sight lines near the bubbles, suggesting the presence of a plasma with $T > 2 \times 10^6$ K.  }

\item{Our best-fit parametric model implies $n_{FB} = 8.2 \pm 0.2 \times 10^{-4}$ cm$^{-3}$, $n_{shell} = 10.0 \pm 0.3 \times 10^{-4}$ cm$^{-3}$, log($T_{FB}$) = 6.60--6.65, and log($T_{shell}$) = 6.60--6.95 with an optimal value of 6.70.  This involves explicitly fitting the \oviii line intensities and analyzing the \oviii/\ovii ratio distribution near the bubbles.  }

\item{These densities imply thermal gas masses within the bubble and shell volumes of $M_{FB}$ = 4.6--5.0$\times 10^6$ $M_{\odot}$ and $M_{shell}$ = 5.2--9.8$\times 10^6$ $M_{\odot}$.  We interpret this as predominantly shock-heated hot gas halo material.  }

\item{The inferred bubble/shell temperature ($5 \times 10^6$ K) compared to ambient halo gas temperature ($2 \times 10^6$ K) suggests a shock Mach number of $\mathcal{M} = 2.3_{-0.4}^{+1.1}$ and expansion rate of $v_{exp} = 490_{-77}^{+230}$ km s$^{-1}$.  These are larger than the values suggested from other soft X-ray modeling analyses \citepalias{kataoka_etal15}, and smaller than the value suggested by the UV absorption line analysis by \citet{fox_etal15}.  The differences are likely explained by geometric assumptions for the latter and modeling the hot gas halo emission for the former.  }

\item{Treating the bubbles as a galactic outflow with Sedov-Taylor expansion solutions leads to an inferred energy injection rate of $2.3_{-0.9}^{+5.1} \times 10^{42}$ erg s$^{-1}$ and age of $4.3_{-1.4}^{+0.8}$ Myr.  These energetics and timescales suggest that the bubbles likely formed from a Sgr A* accretion episode, as opposed to sustained nuclear star formation activity.  }

\item{Our results are broadly consistent with predictions from MHD simulations of galactic outflows.  The constrained energy injection rate and age are most consistent with simulations that generate the bubbles from a relatively weak AGN wind \citep{mou_etal14,mou_etal15}.  }

\end{enumerate}

This analysis is an initial effort to constrain the Fermi bubbles' thermal gas structure using soft X-ray observations, but it should also motivate future observational and theoretical studies.  Future analyses using additional spectral data or all-sky maps from MAXI or eROSITA will help probe the bubbles' structure and interaction with the surrounding medium.  The results should also motivate future simulations that predict characteristic bubble densities, temperatures, pressures, and expansion rates.


\acknowledgments

We thank Meng Su, H.-Y. Karen Yang, Mateusz Ruszkowski, and the anonymous referee for their insightful comments that improved our manuscript.  We also acknowledge the NASA Astrophysics Data Analysis Program grants NNX16AF23G and NNX11AJ55G for funding this work.


\def\apjl{{ApJL}}               
\bibliographystyle{apj}


\end{document}